\documentclass[11pt,cite,epsfig,psfrag]{article}
\usepackage[utf8]{inputenc}
\usepackage{placeins}
\usepackage{geometry}
 \usepackage{graphicx}
 \usepackage{subfig}
  \usepackage{amsmath} 
  \graphicspath{{images@topology_gb/}}
  \usepackage[usenames, dvipsnames]{color}
\usepackage{wrapfig}
\usepackage{boxedminipage}
\usepackage{multirow}

\usepackage{lipsum} 

\usepackage{fullpage}
\usepackage{amsmath}
\usepackage{amssymb}
\usepackage{graphicx}
\usepackage{mathrsfs}
\usepackage{wrapfig}
\usepackage{boxedminipage}
\usepackage{epsfig}
\usepackage{setspace}
\usepackage{float}
\usepackage{hyperref}
\usepackage{enumerate}
\usepackage{cite}
\usepackage[font=footnotesize,labelfont=rm]{caption}

\usepackage[numbers,sort&compress]{natbib}

\def\be{\begin{equation}}
\def\ee{\end{equation}}
\def\bea{\begin{eqnarray}}
\def\eea{\end{eqnarray}}

\numberwithin{equation}{section}

 \newcommand{\RN}[1]{%
   \textup{\uppercase\expandafter{\romannumeral#1}}%
 }
\begin{document}

\thispagestyle{empty}

\vskip 2cm

\begin{center}
{\Large \bf Topology of black hole thermodynamics in Gauss-Bonnet gravity}
\end{center}

\vskip .2cm

\vskip 1.2cm

\centerline{ \bf   Pavan Kumar Yerra \footnote{pk11@iitbbs.ac.in} and Chandrasekhar Bhamidipati\footnote{chandrasekhar@iitbbs.ac.in}
}

\vskip 7mm 
\begin{center}{ School of Basic Sciences\\ 
Indian Institute of Technology Bhubaneswar \\ Bhubaneswar, Odisha, 752050, India}
\end{center}

\vskip 1.2cm
\vskip 1.2cm
\centerline{\bf Abstract}
\vskip 0.5cm
Thermodynamics of black holes in anti de Sitter (AdS) spacetimes typically contains critical points in the phase diagram, some of which correspond to the first order transition ending in a second order one. Following the recent proposal in~\cite{Wei:2021vdx} on using Duan's $\phi$-mapping theory, we classify the critical points of six dimensional charged Gauss-Bonnet black holes in AdS spacetime. We find that the higher derivative curvature terms in the form of Gauss-Bonnet gravity do not change the topological class of critical points in charged black holes in AdS, unlike the case of Born-Infeld corrections noted earlier. The connection between the topological nature of critical points and existence of first order phase transitions breaks down in a certain parameter regime. A resolution is proposed by treating the novel and conventional critical points as phase creation and phase annihilation points, respectively. Examples are provided to support the proposal. 
\noindent

\newpage
\setcounter{footnote}{0}
\noindent

\baselineskip 15pt

\section{Introduction}

Black holes provide a rich arena to explore the effects of strong gravity~\cite{LIGO2017}.
Black hole thermodynamics in particular gives us deep insights to understand the nature of degrees of freedom of gravity~\cite{Sen:2007qy}.  Classically, black holes can be assigned an entropy $S$ which is a constant times the surface area of the even horizon $A$~\cite{Bekenstein:1973ur}, and this leads to the identification of laws of mechanics with the corresponding laws of thermodynamics~\cite{Bardeen:1973gs}. The formal analogy with thermodynamics comes to life only after invoking quantum effects where the temperature $T$ is identified with surface gravity $\kappa$~\cite{Hawking:1975vcx}. Black hole mass $M$ on the other hand is related to internal energy $U$. Hawking and Page discovered a remarkable transition which happens between a self-gravitating hot radiation and Schwarzschild black hole, albeit in asymptotically anti de Sitter (AdS) spacetime~\cite{Hawking:1982dh}. The Hawking-Page (HP) transition of course has an alternate interpretation as a confinement/deconfinement transition in the dual conformal field theory (CFT)~\cite{Maldacena:1997re,Gubser:1998bc,Witten:1998qj}. In the case of charged black holes in AdS, there exists a first-order phase transition between the small and large black holes analogous to liquid-gas phase transitions of van der Waals (VdW) fluid~\cite{Chamblin:1999tk}. The situation becomes interesting in the black hole chemistry paradigm~\cite{Caldarelli:1999xj,Kastor:2009wy,Cvetic:2010jb,Dolan:2011xt,Karch:2015rpa,Kubiznak:2016qmn}, where the cosmological constant $\Lambda$ is assumed to be giving rise to a pressure $P = -\Lambda/8\pi$, with its conjugate thermodynamic volume denoted as $V$. One also realises that the erstwhile first law
\begin{equation}
dM = T dS + \Phi dQ \, ,
\end{equation}
can now include the traditional work terms appearing in standard thermodynamics, when written as~\cite{Kastor:2009wy}
\begin{equation}
dM = T dS + VdP + \Phi dQ  \, ,
\end{equation}
where $Q$ is the charge and $\Phi$ the conjugate potential, with $M$ now reinterpreted as enthalpy $H$. This novel extended thermodynamics ensures  for the charged AdS black holes that, the small-large black hole  transition   has an exact map to the liquid-gas type phase transition, including the presence of a critical region where first order phase transition terminates in a second order one~\cite{Chamblin:1999tk,Caldarelli:1999xj,Kubiznak:2012wp}.  Criticality in phase transitions, especially in black holes is an interesting topic on its own due to several universal phenomena which occur together with the scaling of thermodynamic quantities. Very recently, a remarkable idea was put forward in~\cite{Wei:2021vdx}, where  a topological approach following Duan's $\phi$-mapping theory was used to classify the nature of critical points for charged and Born-Infeld black holes. \\

\noindent
It is interesting as well as important to test whether the proposal~\cite{Wei:2021vdx} is valid just for VdW's liquid-gas type transitions or extendable to more general situations when there are multiple critical points. In fact, apart from the small black hole (SBH) to large black hole (LBH) transition, there are further intriguing phenomena akin to day to day thermodynamic systems, such as reentrant phase transitions, multiple solid-liquid-gas type transitions which occur in a variety of black holes, especially in six or larger dimensions\cite{Altamirano}. The black hole systems involve Gauss-Bonnet, Born-Infeld and other higher derivative curvature terms added to the Einstein action, in addition to multiple rotating Kerr-AdS black hole systems. The later in fact also display tricritical points in certain range of parameters, including the interesting superfluid phenomena\cite{Lu:2010xt,Shen:2005nu,Cvetic:2001bk,Cai:2001dz,AltamiranoKubiznak,Altamirano,Altamirano:2013ane,Cai:2013qga,Kofinas:2006hr,Wei:2019uqg,Hennigar:2016xwd,Banerjee:2011cz}. While there is an independent understanding emerging for several of these phase transitions and critical behaviour from a thermodynamics point of view, it is important to look for other possible methods to make inroads into the classification of such transitions.\\

\noindent
 With the above motivations, we look for a tractable system which presents us with multiple critical points in the phase diagram of black holes, i.e., the Gauss-Bonnet (GB) theory in six dimensions~\cite{Wei:2014hba,Wei:2012ui,Frassino:2014pha}. Higher derivative curvature terms from gravity sector, such as Gauss-Bonnet and Lovelock terms are important, in the discussions of semiclassical quantum gravity, and from the point of view of low-energy effective action of superstring theories. The particular interest in these higher curvature terms comes from the fact that they result in field equations which contain no more than second derivatives of the metric for dimensions higher than five, topological in four and vanishing identically for dimension less than three. Other Lovelock terms behave similarly with respect to their critical dimension
and thus have the advantage of absence of ghost-like modes. In the context of AdS/CFT correspondence, Gauss-Bonnet terms can be understood as providing corrections in the large $N$ expansion of boundary gauge theories, in the strong coupling limit. In the past, such terms have given crucial contributions to the viscosity to entropy ratio and novel bounds~\cite{Brigante:2007nu}.  Although, in general in the Gauss-Bonnet theory, the small to large black hole type phase transition exists in any number of dimensions, the six dimensional case is special. This is because there could either be one, two or even three critical points with presence of multiple swallow tail behaviour, notwithstanding the narrow range of parameters. This still gives us a rich range of parameters to explore the phase diagram and put the proposals in~\cite{Wei:2021vdx}  to test. Another motivation is the surprising result pointed out in~\cite{Wei:2021vdx}, that the charged black holes and Born-Infeld black holes are thermodynamically classified into two different topological classes. This raises the question, as to whether the higher derivative curvature corrections in gravity sector, such as, the Gauss-Bonnet etc., would change the topological nature of critical points of black holes too. \\

\noindent
The rest of the paper is organised as follows. In section-(\ref{phtopo}), we closely follow~\cite{Wei:2014hba,Wei:2012ui,Frassino:2014pha,Wei:2021vdx,Duan:2018rbd,Duan:1984ws} and collect salient aspects of thermodynamics of Gauss-Bonnet AdS black holes in six dimensions and present the topological approach, necessary for performing the calculations in section(\ref{6GB}). 
Section-(\ref{6GB}) contains the main results of calculations on topological classification of critical points of GB black holes in AdS in various charge ranges. Section-(\ref{naturegb}), contains a discussion on the nature of critical points, were we identify a parameter range where the classification of critical points as conventional or novel, proposed in~\cite{Wei:2021vdx} breaks down. We end this section by giving a new proposal on classifying the critical points as phase creation and phase annihilation, to resolve the ambiguity. Remarks and conclusions are given in section-(\ref{conclusions}). Appendix A contains the calculation of topological charge for black holes in six dimensions when the Gauss-Bonnet parameter $\alpha$ is set to zero, which shows that our results are in conformity with the earlier solutions~\cite{Wei:2021vdx} in the Einstein-Maxwell system. In Appendix B, we reason that our novel proposal for understanding topology of critical points presented in section-(\ref{naturegb}), is consistent and works for the case of black holes in third order Lovelock gravity as well.

\section{Thermodynamics of Gauss-Bonnet black holes and topology} \label{phtopo}
We start with the thermodynamics of charged-Gauss-Bonnet (GB) black holes in AdS. As mentioned earlier, this system in six dimensions exhibits a rich phase structure compared to its lower and higher dimensional counter parts~\cite{Wei:2014hba,Wei:2012ui,Frassino:2014pha}. The corresponding thermodynamic quantities, i.e., the temperature $T$, entropy $S$, specific heat $C_P$, and the Gibbs free energy $G$ in the extended phase space (where we treat the cosmological constant $\Lambda$
as the pressure $P$ and its conjugate quantity as thermodynamic volume $V$~\cite{Kastor:2009wy}), are given by~\cite{Wei:2014hba,Frassino:2014pha}:
\begin{eqnarray}
T &=& \frac{8\pi P r_h^8 + 6r_h^6 + 2\alpha r_h^4 -Q^2}{8\pi r_h^5 (r_h^2 + 2\alpha)}, \label{eq:GB_eq of st} \\
S &=& \frac{r_h^4}{4} \Big(1+\frac{4\alpha}{r_h^2}\Big), \label{eq:GB_entropy} \\
C_P &=& \frac{(r_h^3+2 \alpha r_h)^2 (8\pi P r_h^8 + 6r_h^6+2\alpha r_h^4 -Q^2)}{Q^2(7r_h^2 +10\alpha) + 2r_h^4 (4\pi P r_h^6 +3r_h^4(8\pi P \alpha -1) +3\alpha r_h^2 -2\alpha^2)}, \\
G &=& \frac{5Q^2(7r_h^2+20)-6r_h^4(4\pi P r_h^6 +r_h^4(48\pi P -5)+5r_h^2-20)}{480\pi r_h^3 (r_h^2+2)},
\end{eqnarray}
where, $r_h$ is the horizon radius, $Q$ is the charge of the black hole and $\alpha$ is the Gauss-Bonnet (GB) coupling parameter.

\subsection{Phase Structure } \label{phasegb}
\begin{figure}[h!]
 	{\centering
 		\includegraphics[width=3in]{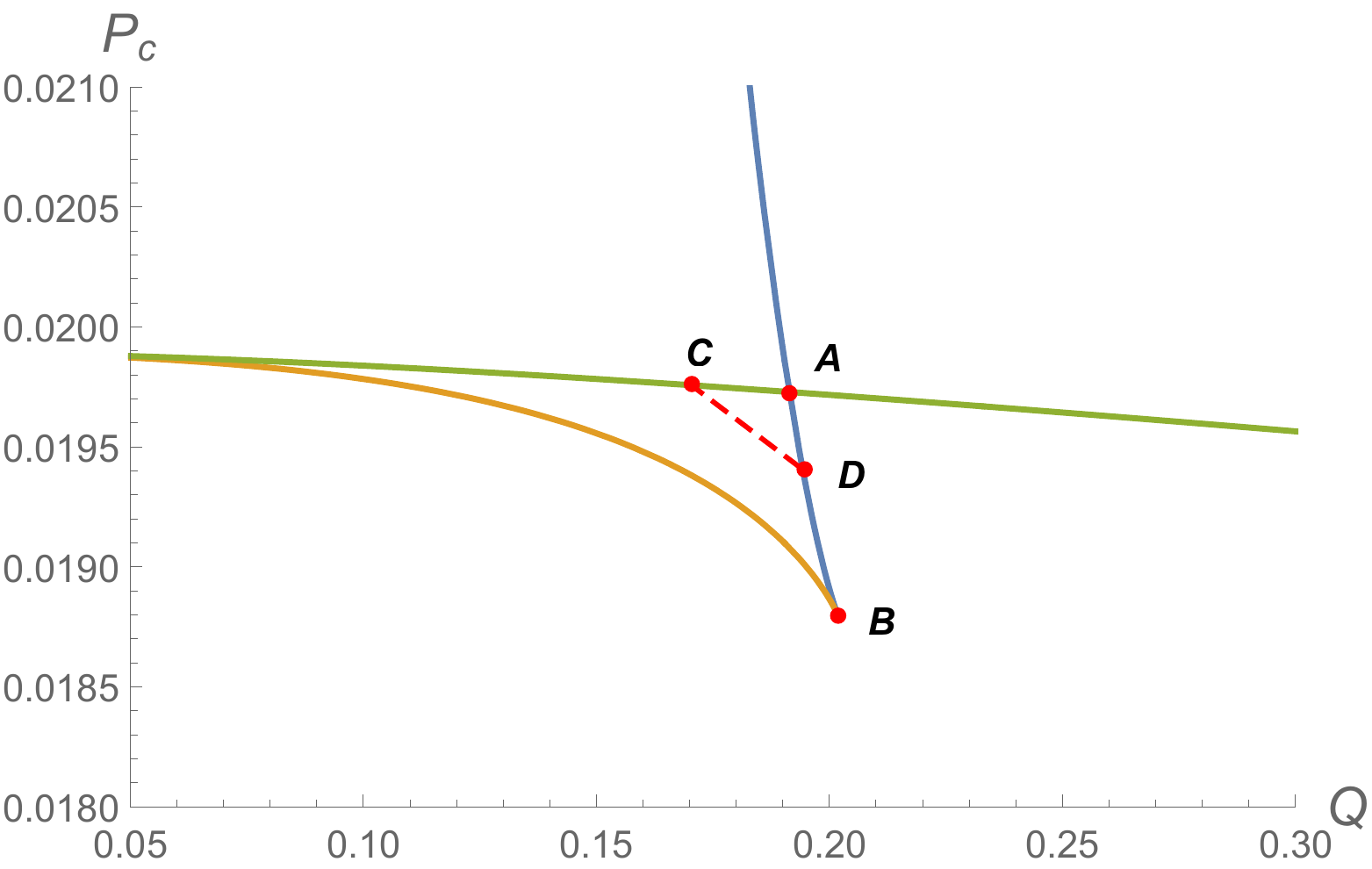}	
 		
 		\caption{\footnotesize Behaviour of critical pressure $P_c$ with charge $Q$. The charges at the points A, B, C, and D are $ Q_A =0.1914, Q_B=0.2018, Q_C=0.1705$, and $Q_D=0.1946$ respectively. We have three critical points for $Q< Q_B$, and one critical point for $Q>Q_B$. The line segment CD denotes the  triple points. (Here, we set $\alpha$ = 1). }\label{fig:pqplot} 
 	}
 \end{figure}

The equation of state (obtained by rewriting eqn. \ref{eq:GB_eq of st} in terms of pressure) exhibits various phase structures depending on the charge parameter $Q$~\cite{Wei:2014hba,Frassino:2014pha}. This can be seen from the behavior of critical\footnote{The expression for $P_c$ can be find in~\cite{Frassino:2014pha,Wei:2014hba}.} pressure $P_c$ with charge $Q$, as shown in Fig.~\ref{fig:pqplot}.
\noindent
 Behaviour of critical pressure $P_c$ shows the existence of three critical points for $Q< Q_B$, two critical points for $Q=Q_B$, and one critical point for  $Q > Q_B$. However, for case $Q=Q_B$, the critical point at $B$ in Fig.~\ref{fig:pqplot}  is not a true critical point, as it is not an inflection point but an undulation point. Thus, we have only one critical point for $Q=Q_B$.
\vskip 0.3cm \noindent
Further, in the case $Q \geq Q_B$, the critical points are  related  to the van der Waals type small/large black hole phase transitions, whereas in the case $Q < Q_B$, the critical points are related to a rich phase structure involving the triple points (where, the three phases, i.e., small/intermediate/large black hole phases, can coexist).  In fact, all the critical points may not globally minimise the Gibbs free energy, and thus may not appear in the phase diagram. The nature of the critical points actually depends on the fixed charge ranges, viz. $ Q<Q_C, \, Q_C < Q< Q_D, \, Q_D <Q<Q_B,$ and  $Q_B < Q$.  
For the detailed discussion of phase structures related to these critical and triple points, one can refer~\cite{Wei:2014hba,Frassino:2014pha}. Our aim in the next subsection, is to present the concept of topology in thermodynamics, which can be used to study properties associated with these critical points that belong to different fixed charge ranges. 

\subsection{Topology of thermodynamical functions} \label{topology}
Recently, it is shown in~\cite{Wei:2021vdx} that, the critical points  in the phase diagram of a  thermodynamic system (black holes in particular) could be classified into conventional and novel  critical type, based on the topological charges they carry. 
In the extended thermodynamic framework, typically the temperature $T$ of a thermodynamic system is given
 as a function of the entropy $S$, pressure $P$, and other parameters $x^i$, i.e.,
 \begin{equation}\label{eq:general_T}
 T =  T(S, P, x^i) \, .
 \end{equation}  
It is well known that the critical point, where the second order transition happens in the phase diagram, can be obtained by solving the condition for stationary point of inflection, i.e., 
\begin{equation}\label{eq:inflection}
(\partial_S T )_{P, x^i} =0, \ \ (\partial_{S,S} T )_{P, x^i} =0 \, .
\end{equation}
\noindent 
Now, the interesting suggestion put forward in~\cite{Wei:2021vdx} is to construct a scalar thermodynamic function as
\begin{equation}
\Phi = \frac{1}{\text{sin} \theta} \, T(S, x^i) \, .
\end{equation}
In practice, this is found by eliminating one of the variables in eqn. (\ref{eq:general_T}), through the first of the conditions in eqn. (\ref{eq:inflection}), and adding an additional factor of $1/\sin\theta$ for the ease of analysis. The set up of Duan's $\phi$-mapping theory proceeds by constructing a new vector field 
$\phi = (\phi^S, \phi^\theta)$, where $\phi^S = (\partial_S \Phi)_{\theta, x^i}$ and $\phi^\theta = (\partial_\theta \Phi)_{S, x^i}$.  $\phi $ has an important property that, its zero points always lie at $\theta =  \pi/2$,  and can be identified with the presence of the critical points of the thermodynamic system. The horizontal lines at $\theta = 0$ and  $\pi$, act as the  boundaries of the parameter space, where the vector field $\phi$ is perpendicular to these lines.  An important property of the above construction is the presence of a topological current $j^{\mu}$, whose non-zero contribution only comes from the zero points of the vector field $\phi^a$, i.e., $\phi^a(x^i) = 0$. Let there be $N$ solutions of $\phi^a$, whose $i$-th solution is denoted as $\vec{x} =\vec{z_i}.$ $Q_t =\int_\Sigma j^0 d^2x$ thus gives the corresponding charge, which can be computed as~\cite{Duan:1984ws,Duan:2018rbd,Wei:2021vdx}
\begin{eqnarray}\label{tcharge}
Q_t &=& \int_\Sigma \Sigma_{i=1}^N \beta_i  \eta_i  \delta^2(\vec{x}-\vec{z_i})d^2x  \nonumber \\ 
 &=& \Sigma_{i=1}^N \beta_i \eta_i = \Sigma_{i=1}^N w_i.
\end{eqnarray}
Here, $\beta_i$ is the positive integer (Hopf index) measuring the number of loops that $\phi^a$ makes around the $i$-th zero point of $\phi$.  $\eta_i = \text{sign}(J^0(\phi/x)_{z_i}) = \pm 1$ is called the Brouwer degree and $w_i$ is the winding number for $i$-th zero point of $\phi$. Since, $Q_t$ is non-zero only at the zero points of $\phi$, one can assign a topological charge (given by the winding number) for each critical point, where the vector field $\phi$ is zero. 
Since, the Brouwer degree $\eta_i$ can be positive or negative, the critical points were proposed to be further divided into two different topological classes, i.e., the conventional (where $\eta_i = -1$) and the novel  (where $\eta_i = +1$)~\cite{Wei:2021vdx}. Further proposal is that, allowing $\Sigma$ to span the entire parameter space for a given thermodynamic system, the topological properties of different thermodynamic systems can be divided into different classes from thermodynamics point of view.

\section{Topology of critical points in six dimensional charged Gauss-Bonnet black holes in AdS} \label{6GB}

First we obtain the general expression for topological charge which can be used to analyse the critical points of black holes in six dimensional charged-Gauss-Bonnet theories in AdS spacetime.
\subsection{Topological Charge} \label{topologygb}
Starting from the temperature~\footnote{Here, for a given $\alpha$, entropy $S=S(r_h)$ from eqn.~\eqref{eq:GB_entropy}.} $T$ in eqn.~\eqref{eq:GB_eq of st} and following the discussion in section-(\ref{topology}), we now compute the thermodynamic function $\Phi$, as 
\begin{eqnarray}
\Phi &=& \frac{1}{\text{sin}\theta} T(r_h, Q, \alpha), \\
&=& \frac{1}{\text{sin}\theta} \frac{(3r_h^6 + 2r_h^4 \alpha -2Q^2)}{2\pi r_h^5 (r_h^2 + 6\alpha)}.
\end{eqnarray}
The vector field $\phi= (\phi^r, \phi^\theta)$ is obtained to be
\begin{eqnarray}
\phi^r &=& \partial_{r_h} \Phi = \frac{\text{csc}\theta \Big(2Q^2(7r_h^2+30\alpha) - 3r_h^4(r_h^2-2\alpha)^2\Big)}{2\pi r_h^6 (r_h^2+6\alpha)^2}, \\
\phi^\theta &=& \partial_\theta \Phi = -\frac{\text{cot}\theta \, \text{csc}\theta (3r_h^6+2r_h^4\alpha -2Q^2)}{2\pi r_h^5 (r_h^2+6\alpha)}. 
\end{eqnarray}
The normalised vector field is thus $n=(\frac{\phi^r}{||\phi||},\frac{\phi^\theta}{||\phi||})$.
In order to calculate the topological charge associated with a critical point (where $\phi = 0$), one is required to find its winding number $w_i$. We know from the topology that if a given contour encloses a critical point then its winding number (i.e., the topological charge) is non-zero, otherwise it is zero~\cite{Wei:2021vdx,Cunha:2020azh,Junior:2021svb,Wei:2020rbh,Cunha:2017qtt}.
\vskip 0.3cm \noindent
In the  orthogonal $\theta - r$ plane, we consider a contour $C$,  that is piece-wise smooth and positive oriented. Let the contour $C$, for simplicity, be an ellipse centered at $(r_0, \frac{\pi}{2})$, parameterised by the angle $\vartheta \in (0, 2\pi)$ as~\cite{Wei:2021vdx,Wei:2020rbh}:
\begin{eqnarray}
\left\{
\begin{aligned}
r&=a\cos\vartheta+r_0, \\
\theta&=b\sin\vartheta+\frac{\pi}{2}.
\end{aligned}
\right.
\end{eqnarray}
Then, following~\cite{Wei:2021vdx,Junior:2021svb,Wei:2020rbh,Cunha:2020azh}, one can compute the topological charge (i.e., winding number) by measuring the deflection $\Omega(\vartheta)$ of the vector field $\phi$ along the given contour as
\begin{equation}
Q_t = \frac{1}{2\pi} \Omega (2\pi),
\end{equation}
where,
\begin{equation}\label{eq:deflection}
\Omega(\vartheta)=\int_{0}^{\vartheta}\epsilon_{ab}n^{a}\partial_{\vartheta}n^{b}d\vartheta.
\end{equation}
\vskip 0.3cm \noindent 
In the following subsections, we compute the topological charges corresponding to the critical points of various fixed charge ranges given in Table~\ref{table:coeff}  and Table~\ref{table:cri}. {\bf Case 1} in the Table~\ref{table:coeff} below corresponds to the Einstein-Maxwell case with a single critical point, where the Gauss-Bonnet parameter $\alpha=0$. The calculation of topological charge for this case is given in Appendix-(\ref{AppendixA}) and shows that the classification of critical points matches the earlier considerations in lower dimensions~\cite{Wei:2021vdx}.
\vskip 0.3cm \noindent 
\begin{table}[!htb]
	
	\begin{minipage}{.5\linewidth}
	\centering{	
		\begin{tabular}{|c|c|c|c|c|c|c|}
			\hline \hline 
			\multicolumn{2}{|c|}{Case} &  $C_1$ & $C_2$ & $C_3$ & $C_4$ & $C_5$ \\ \hline  
			\multirow{3}{4em} {Case-1} & a & $0.07$ & $0.07$ & - & - & - \\ 
			& b & $0.4$ & $0.4$ & - & - & - \\ 
			& $r_0$ & $0.69$ & $0.9$ & - & - & - \\ \hline
			\multirow{3}{4em} {Case-2} & a & 0.15 & 0.15 & 0.15 & 0.15 & 0.7 \\ 
			& b & 0.2 & 0.2 & 0.2 & 0.2 & 0.5 \\ 
			& $r_0$ & 1.2 & 1.55 & 0.67 & 2.2 & 1.1 \\ \hline
			\multirow{3}{4em} {Case-3} & a & 0.15 & 0.15 & 0.15 & 0.6 & - \\ 
			& b & 0.4 & 0.4 & 0.4 & 0.7 & - \\ 
			& $r_0$ & 1.15 & 1.55 & 0.8 & 1.2 & - \\ \hline
			\multirow{3}{4em} {Case-4} & a & 0.1 & 0.1 & 0.1 & 0.6  & - \\ 
			& b & 0.4 & 0.4 & 0.4 & 0.8 & - \\ 
			& $r_0$ & 1.1 & 0.87  & 1.55 & 1.2 & - \\ \hline
			\multirow{3}{4em} {Case-5} & a & 0.2 & 0.2 & - & - & - \\ 
			& b & 0.4 & 0.4 & -& - & -  \\ 
			& $r_0$ & 1.61 & 2.2 & - & - & -  \\
			\hline\hline 
		\end{tabular}
		\caption{Parametric coefficients of contours.}
		\label{table:coeff} }
	\end{minipage} \hskip 0.2cm	
	\begin{minipage}{.5\linewidth}
		\centering{		 
			\begin{tabular}{|c|c|c|c|c|c|c|}
				\hline \hline
 \multicolumn{2}{|c|}{Case}  & \multicolumn{3}{|c|}{Critical Point (CP)}    \\ \cline{3-5}
 \multicolumn{2}{|c|}{} & $\text{CP}_1$ & $\text{CP}_2$ & $\text{CP}_3$ \\ \hline
				\multirow{3}{3em} {Case-2} & $r_c$ & 1.21313 & 1.53448 & 0.676405  \\ 
				& $P_c$ & 0.0195571 & 0.0197829 & 0.0321348  \\ 
				& $T_c$ & 0.11228 & 0.112424 & 0.115055 \\ \hline
				\multirow{3}{3em} {Case-3} & $r_c$ & 1.13426 & 1.55453 & 0.792783 \\ 
				& $P_c$ &  0.0192661 & 0.0197442 & 0.021548 \\ 
				& $T_c$ &  0.11208 & 0.112381 & 0.11271 \\ \hline
				\multirow{3}{3em} {Case-4} & $r_c$ & 1.06666 & 0.87761 & 1.5642  \\ 
				& $P_c$ & 0.0190022 & 0.0193334 & 0.0197237 \\ 
				& $T_c$ & 0.111932 & 0.11203 & 0.11236\\ \hline
				\multirow{3}{3em} {Case-5} & $r_c$ & 1.62636 &- & - \\ 
				& $P_c$ & 0.0195648 & - & - \\ 
				& $T_c$ & 0.11217 & - & - \\
				\hline\hline
			\end{tabular}
			\caption{Critical values.}
			\label{table:cri}	}
	\end{minipage} 
\end{table}

\subsection{Case 2: \bf $ Q < Q_C $} \label{case:2}
\begin{figure}[h!]
	{\centering
		\subfloat[]{\includegraphics[width=3in]{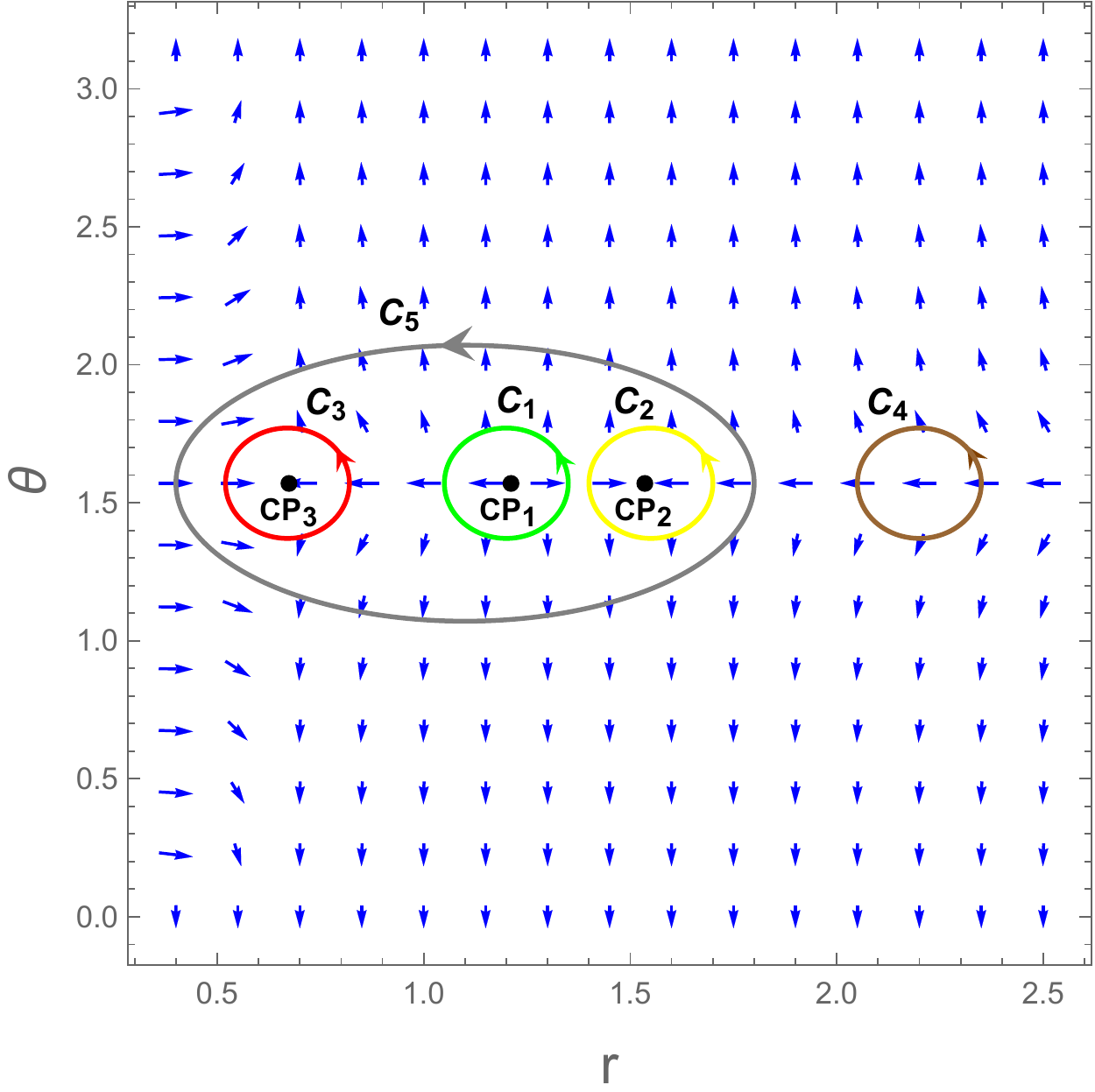}\label{Fig:vecplot_q15}}\hspace{0.5cm}	
		\subfloat[]{\includegraphics[width=3in]{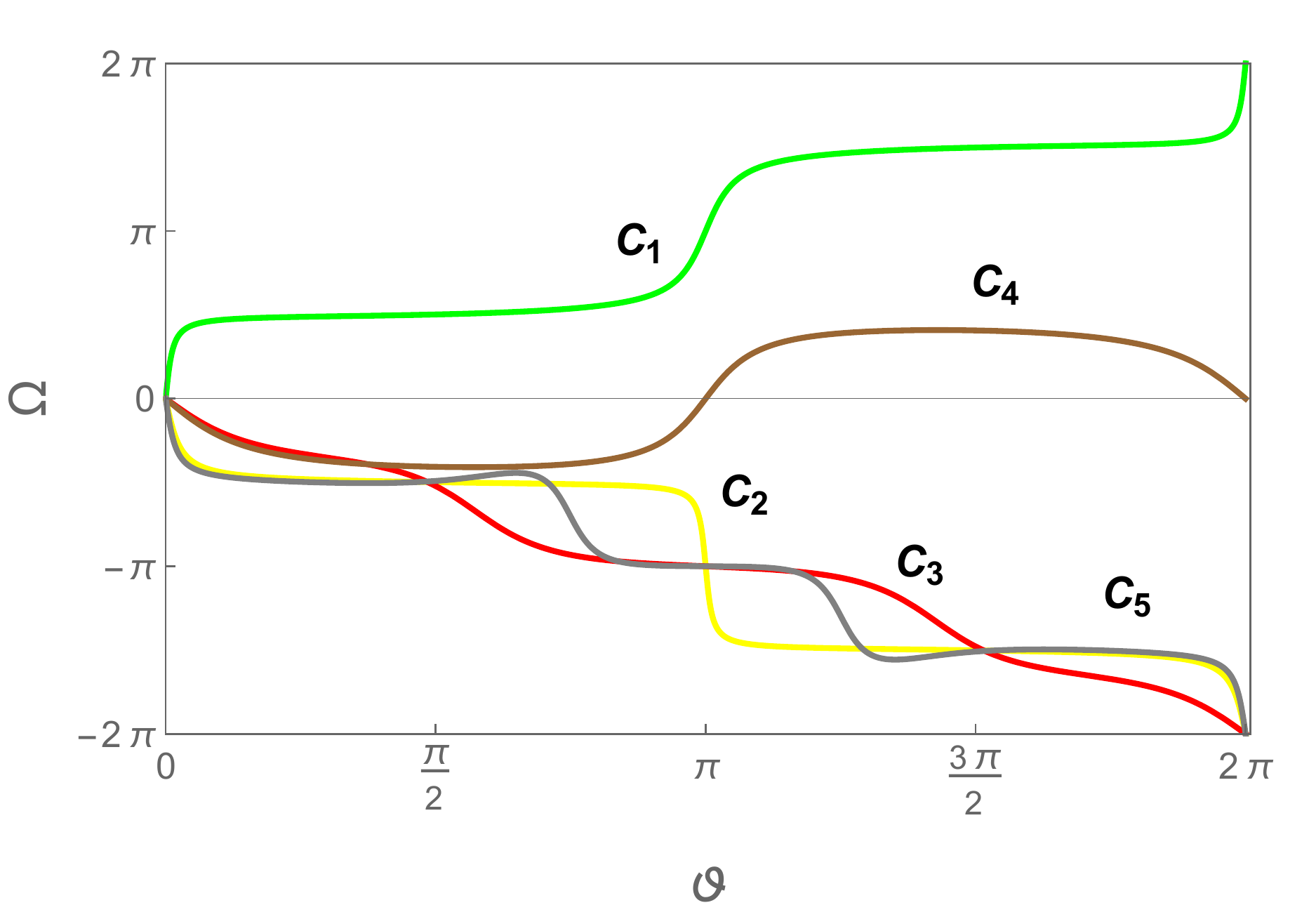}\label{Fig:omegaplot_q15}}				
		
		\caption{\footnotesize For case-2: (a) The blue arrows represent the vector field $n$ on a portion of the $\theta-r$ plane. The critical points $\text{CP}_1$, $\text{CP}_2$, and $\text{CP}_3$ are located at $(r,\theta)=(1.21, \frac{\pi}{2}), (1.53, \frac{\pi}{2}), \, \text{and} \, (0.67, \frac{\pi}{2}) $ marked with  black dots, and they are enclosed with the contours $C_1$, $C_2$ and $C_3$, respectively. 
			The  contour $C_4$  does not enclose any critical point, while the contour $C_5$ encloses all the three critical points. (b) $\Omega$ vs $\vartheta$ for contours $C_1$ (green curve),  $C_2$ (yellow curve), $C_3$ (red curve), $C_4$ (brown curve), and $C_5$ (grey curve).} 
	}
\end{figure}

For simplicity, now onwards we set the Gauss-Bonnet parameter $\alpha=1$.
In this case, we have three critical points (see Table~\ref{table:cri} for the critical values at $Q=0.15$) related to rich phase structure of small/large black hole phase transitions and the Gibbs free energy also exhibits the double swallow tail behaviour~\cite{Wei:2014hba,Frassino:2014pha}. 
For this case, the vector field $n$ is plotted in Fig.~\ref{Fig:vecplot_q15}, where it clearly shows the presence of three critical points.  
We construct five contours  $C_1$, $C_2,$  $C_3$, $C_4$, and $C_5$, such that the contours $C_1$, $C_2,$ and $C_3$ enclose the critical points $\text{CP}_1$, $\text{CP}_2$, and $\text{CP}_3$, respectively, while the contour $C_4$ does not enclose any critical point. The contour $C_5$ encloses all the three critical points (see the Table~\ref{table:coeff} for parametric coefficients of the contours.). 
\vskip 0.3cm \noindent
The behaviour of the deflection angle $\Omega(\vartheta)$ for these contours is as shown in Fig~\ref{Fig:omegaplot_q15}, which gives $\Omega(2\pi) = 2\pi, -2\pi, -2\pi, 0, \, \text{and} \, -2\pi$, for the contours $C_1, C_2, C_3, C_4$, and $C_5$, respectively.
Then, the topological charges associated with the critical points are given by $Q_t|_{\text{CP}_1} = +1, \, Q_t|_{\text{CP}_2} = -1$, and $Q_t|_{\text{CP}_3} = -1$. Again the topological charge for the contour $C_4$ is zero as it  does not enclose the critical point, whereas the topological charge for the contour $C_5$ is $-1$, as it encloses all the three critical points and since its topological charge is additive.  
\vskip 0.3cm \noindent
Again, following the proposal in~\cite{Wei:2021vdx}, $\text{CP}_1$ is now a novel critical point, while  $\text{CP}_2$, and  $\text{CP}_3$, are conventional  critical points. Therefore, in this case, the total topological charge of the system is $Q_t = Q_t|_{\text{CP}_1} + Q_t|_{\text{CP}_2}+Q_t|_{\text{CP}_3} = -1.$

\subsection{Case 3: \bf $Q_C < Q < Q_D$}
\begin{figure}[h!]
	{\centering
		\subfloat[]{\includegraphics[width=3in]{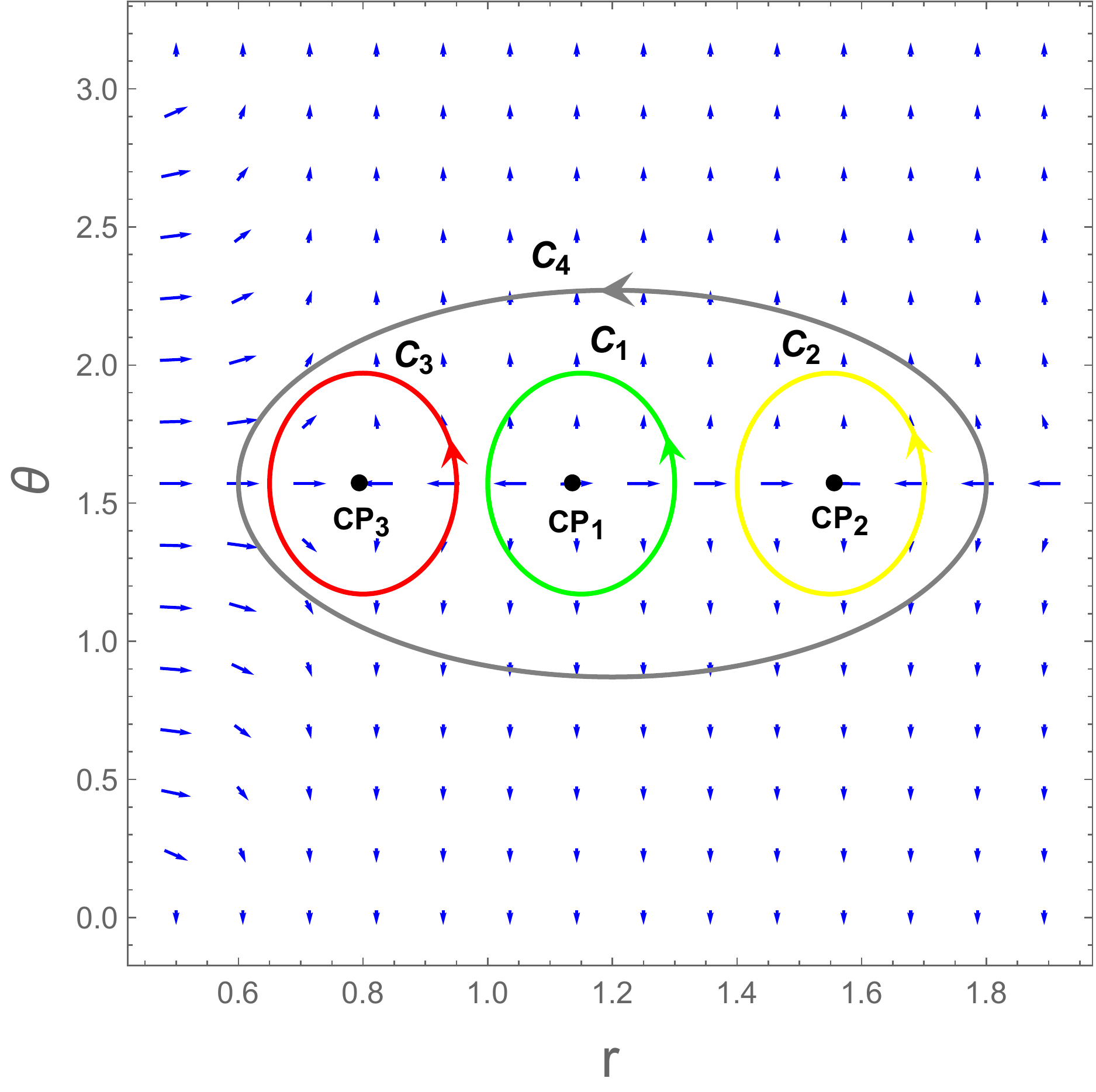}\label{Fig:vecplot_q18}}\hspace{0.5cm}	
		\subfloat[]{\includegraphics[width=3in]{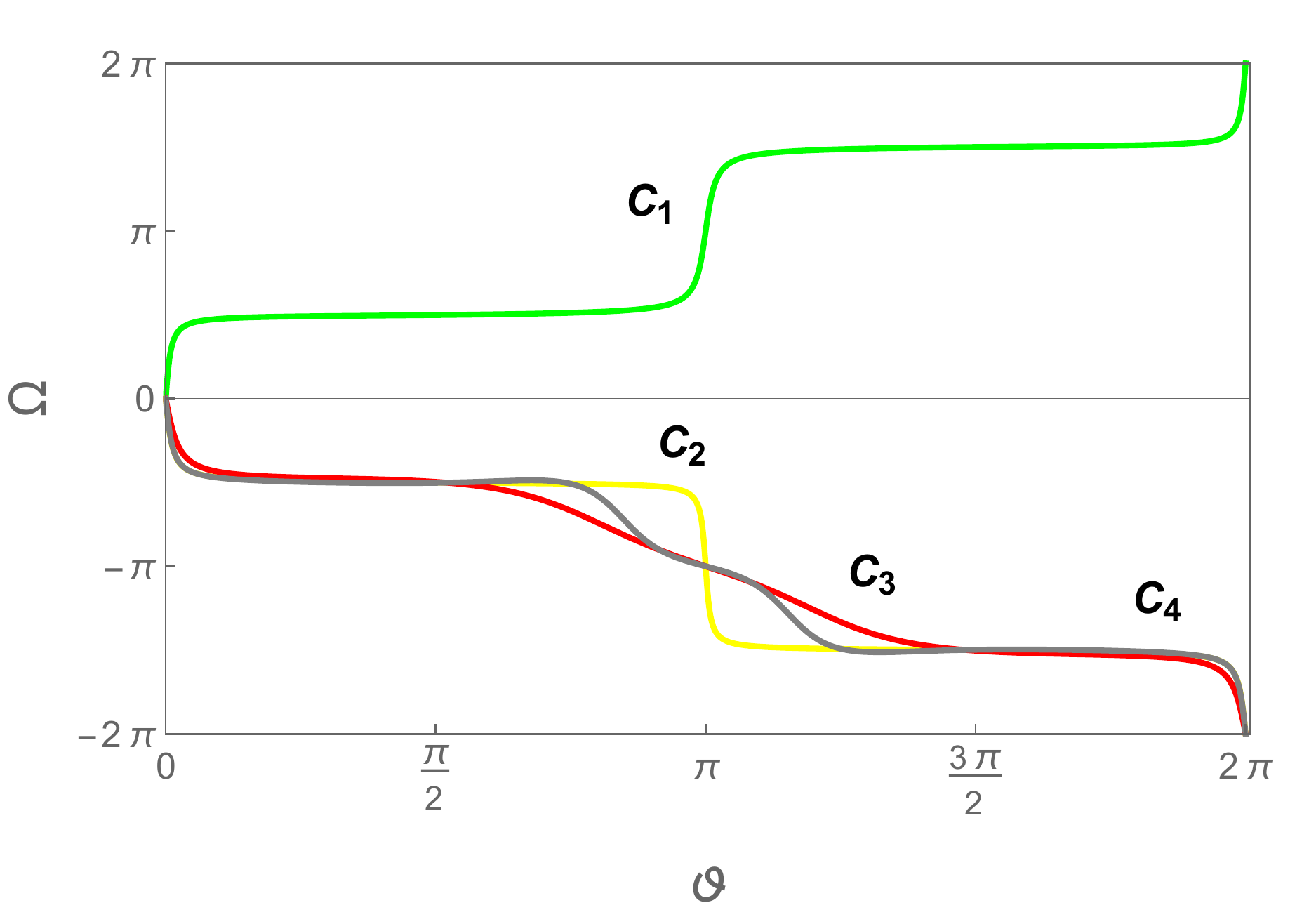}\label{Fig:omegaplot_q18}}				

		\caption{\footnotesize For case-3: (a) The blue arrows represent the vector field $n$ on a portion of the $\theta-r$ plane. The critical points $\text{CP}_1$, $\text{CP}_2$, and $\text{CP}_3$ are located at $(r,\theta)=(1.13, \frac{\pi}{2}), (1.55, \frac{\pi}{2}), \, \text{and} \, (0.79, \frac{\pi}{2}) $ marked with  black dots, and they are enclosed with the contours $C_1$, $C_2$ and $C_3$, respectively. 
			The  contour $C_4$   encloses all the three critical points. (b) $\Omega$ vs $\vartheta$ for contours $C_1$ (green curve),  $C_2$ (yellow curve), $C_3$ (red curve), and $C_4$ (grey curve).} 
	}
\end{figure}

In this case, we have three critical points which give rise to a nice phase structure consisting of small/intermediate/large black hole phase transitions, including the appearance of triple~\footnote{Here, we do not pursue the case of triple points.} points~\cite{Wei:2014hba,Frassino:2014pha}.
The vector field $n$ plotted in Fig.~\ref{Fig:vecplot_q18} shows the three critical points (see Table~\ref{table:cri} for critical values at charge $Q=0.18$).
\vskip 0.3cm \noindent
We construct four contours $C_1$, $C_2$, $C_3$, and $C_4$, such that the contours $C_1$, $C_2$, and $C_3$ enclose the critical points $\text{CP}_1$, $\text{CP}_2$, and $\text{CP}_3$, respectively, while the contour $C_4$ encloses all the three critical points. The behaviour of the deflection angle $\Omega(\vartheta)$ along  these contours is shown in Fig~\ref{Fig:omegaplot_q18}, from where we obtain that $\Omega(2\pi) = 2\pi, \, -2\pi, \, -2\pi, \, -2\pi$, for the contours $C_1$, $C_2$, $C_3$, and $C_4$, respectively.
Then,  the topological charges corresponding to the critical points are $Q_t|_{\text{CP}_1} = +1$, $Q_t|_{\text{CP}_2} = -1$, and $Q_t|_{\text{CP}_3} = -1$. Thus, the critical point $\text{CP}_1$ is a novel one, while the critical points  $\text{CP}_2$, and  $\text{CP}_3$, are conventional critical points.  
\vskip 0.3cm \noindent
Therefore, in this case also, the total topological charge of the system (given by the contour $C_4$) is $Q_t = -1$.
\subsection{Case 4: \bf  $ Q_D < Q < Q_B$}
\begin{figure}[h!]
	{\centering
		\subfloat[]{\includegraphics[width=3in]{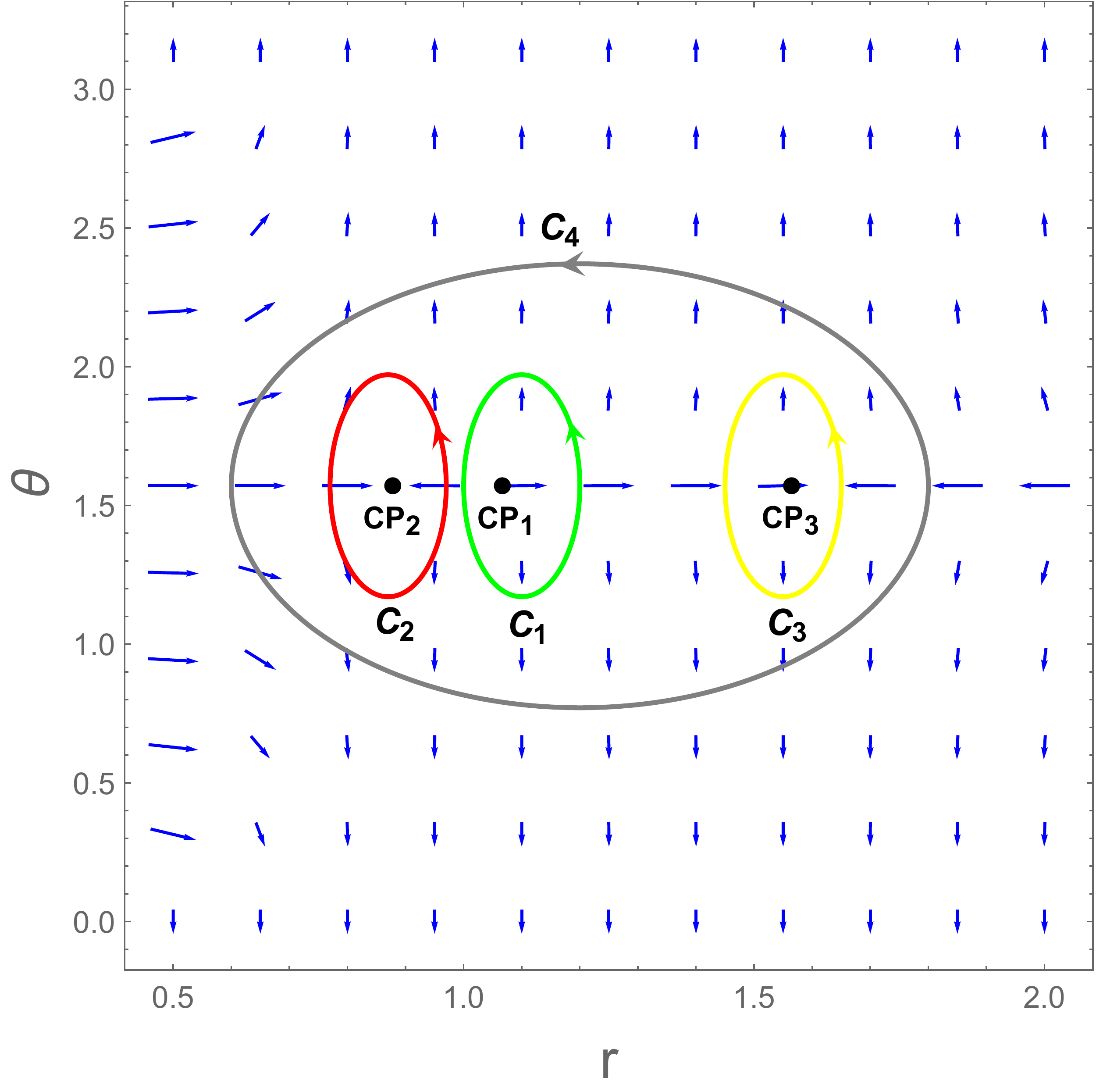}\label{Fig:vecplot_q195}}\hspace{0.5cm}	
		\subfloat[]{\includegraphics[width=3in]{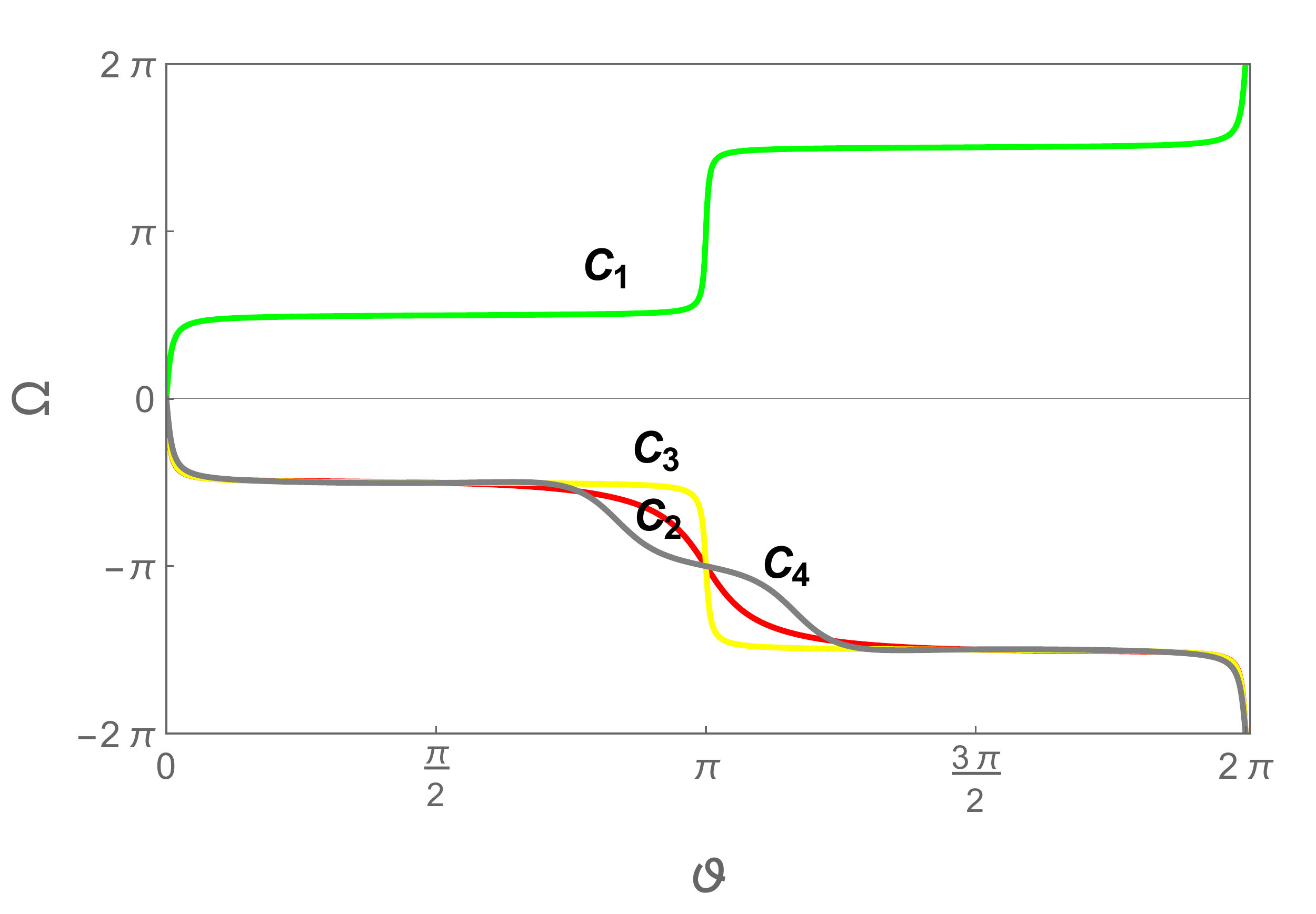}\label{Fig:omegaplot_q195}}				
		
		\caption{\footnotesize For case-4: (a) The blue arrows represent the vector field $n$ on a portion of the $\theta-r$ plane. The critical points $\text{CP}_1$, $\text{CP}_2$, and $\text{CP}_3$ are located at $(r,\theta)=(1.06, \frac{\pi}{2}), (0.87, \frac{\pi}{2}), \, \text{and} \, (1.56, \frac{\pi}{2}) $ marked with  black dots, and they are enclosed with the contours $C_1$, $C_2$ and $C_3$, respectively. 
			The  contour $C_4$   encloses all the three critical points. (b) $\Omega$ vs $\vartheta$ for contours $C_1$ (green curve),  $C_2$ (red curve), $C_3$ (yellow curve), and $C_4$ (grey curve).} 
	}
\end{figure}

In this case also we have three critical points with the existence of  small/large black hole phase transitions~\cite{Wei:2014hba,Frassino:2014pha}. The vector field $n$,  showing these critical points (see Table~\ref{table:cri} for critical values at charge $Q=0.195$), is plotted in Fig~\ref{Fig:vecplot_q195}.
\vskip 0.3cm \noindent
The critical points $\text{CP}_1$, $\text{CP}_2$, and $\text{CP}_3$, are enclosed by the contours $C_1$, $C_2$, and $C_3$, respectively, and the contour $C_4$  encloses all the three critical points.
For these contours $C_1$, $C_2$, $C_3$, and $C_4$, the deflection angle $\Omega(\vartheta)$ behaviour is as shown in Fig.~\ref{Fig:omegaplot_q195}, from where we note that $\Omega(2\pi) = +2\pi, \, -2\pi, \, -2\pi,$ and $-2\pi$, respectively. 
Then, the corresponding topological charges of the critical points are $Q_t|_{\text{CP}_1} = +1$ (novel), $Q_t|_{\text{CP}_2} = -1$ (conventional), and $Q_t|_{\text{CP}_3} = -1$ (conventional). 
\vskip 0.3cm \noindent
The total topological charge of the system is $Q_t = -1$, as in the previous case.  

\subsection{Case 5: \bf $Q_B < Q$}
In this case, there is only one critical point. The situation is similar to the van der Waals system, where the first order small/large black hole phase transition which terminates at this second order critical point~\cite{Wei:2014hba,Frassino:2014pha}.

\begin{figure}[h!]
	{\centering
		\subfloat[]{\includegraphics[width=3in]{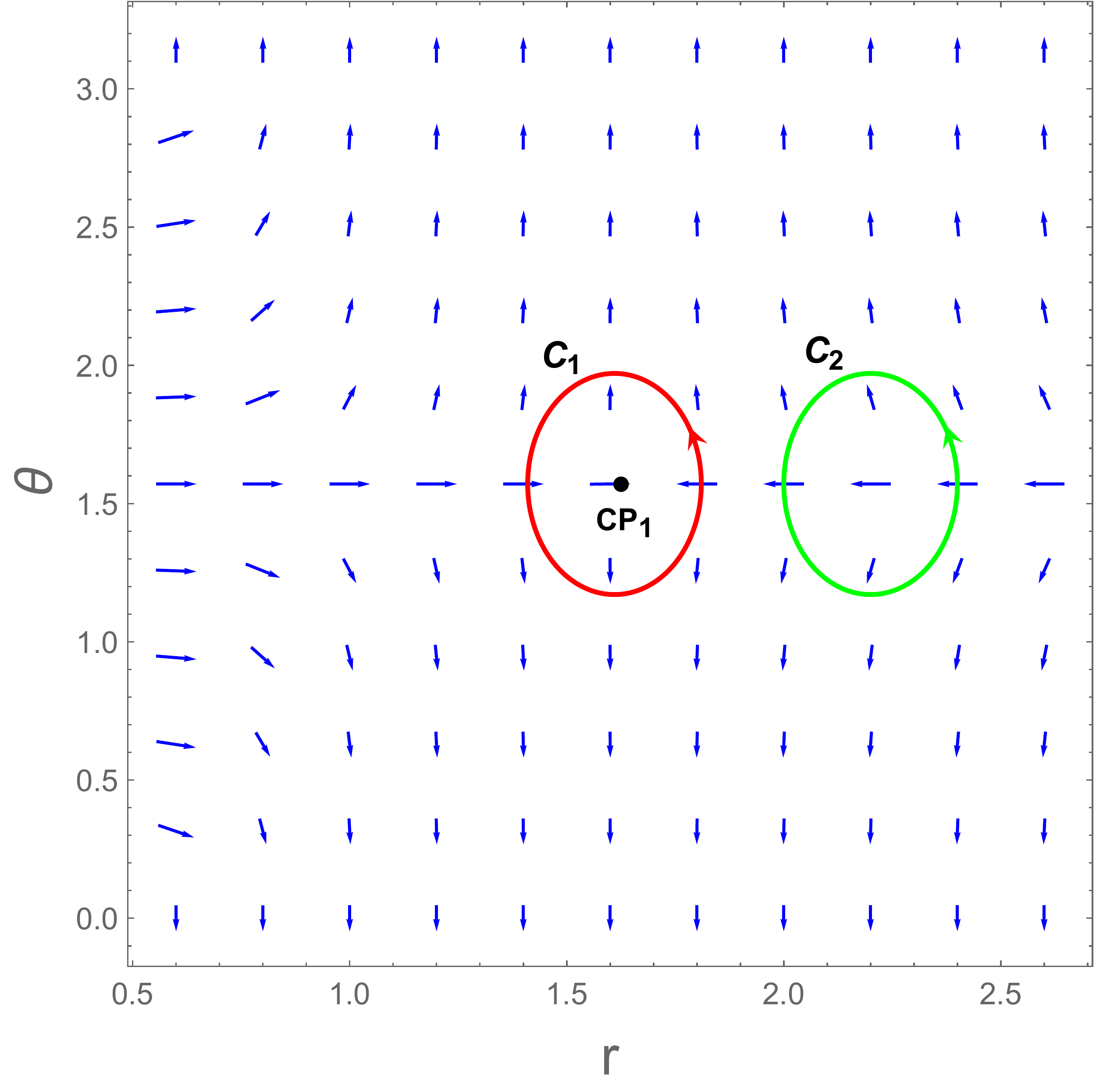}\label{Fig:vecplot_q3xtra}}\hspace{0.5cm}	
		\subfloat[]{\includegraphics[width=3in]{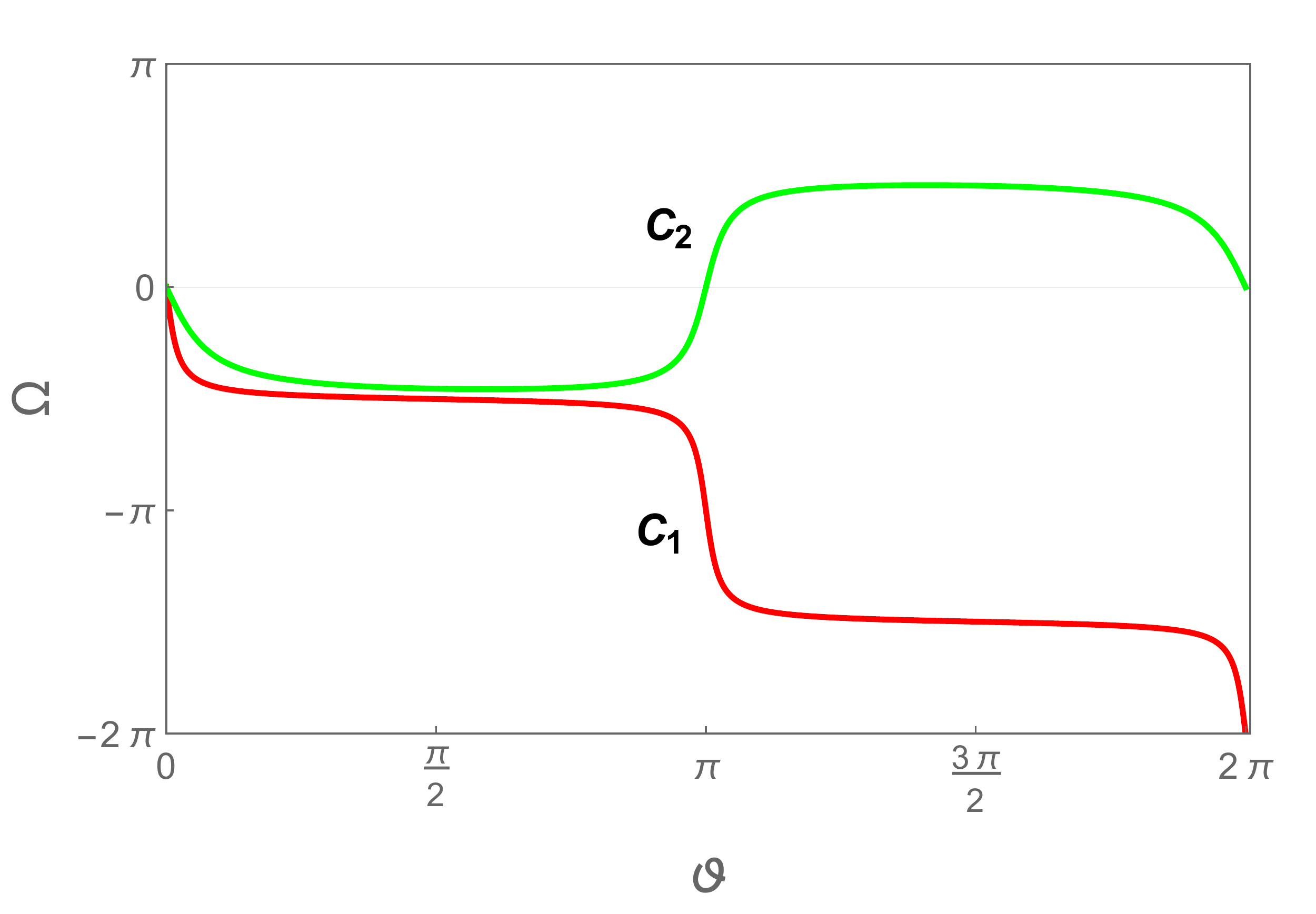}\label{Fig:omegaplot_q3xtra}}				
		
		\caption{\footnotesize For case-5: (a) The blue arrows represent the vector field $n$ on a portion of the $\theta-r$ plane. The critical point $\text{CP}_1$ is located at $(r,\theta)=(1.63, \frac{\pi}{2})$ marked with  black dot, and is enclosed with the contour $C_1$. 
			The  contour $C_2$  does not enclose the  critical point. (b) $\Omega$ vs $\vartheta$ for contours $C_1$ (red curve), and $C_2$ (green curve).} 
	}
\end{figure}

\vskip 0.3cm \noindent
For this case (see the Table~\ref{table:cri} for critical values at charge $Q=0.3$), the vector field $n$ is plotted  Fig.~\ref{Fig:vecplot_q3xtra},  showing the  critical point $\text{CP}_1$. We construct two contours $C_1$, and $C_2$, such that the contour $C_1$ encloses the critical point, while the contour $C_2$ does not. The topological charge  must be non-zero for the contour $C_1$ and zero for the contour $C_2$.
The plot of the deflection angle $\Omega(\vartheta)$ for the contours $C_1$, and $C_2$ , is as shown in Fig.~\ref{Fig:omegaplot_q3xtra}. The angle  $\Omega(\vartheta)$ decreases to reach $-2\pi$ for the contour $C_1$, whereas it decreases first, then increases and finally vanishes for the contour $C_2$. Thus, as we expected, the topological charge for the contour $C_1$ is $-1$, and for the contour $C_2$ it is zero.  
\vskip 0.3cm \noindent
Therefore, in this case, the topological charge of the critical point $\text{CP}_1$ is $Q_t|_{\text{CP}_1} = -1$ (conventional critical point), which is the total topological charge of the system as well.

\section{Nature of the Critical points} \label{naturegb}
\begin{figure}[h!]
	{\centering
		\subfloat[]{\includegraphics[width=2.9in]{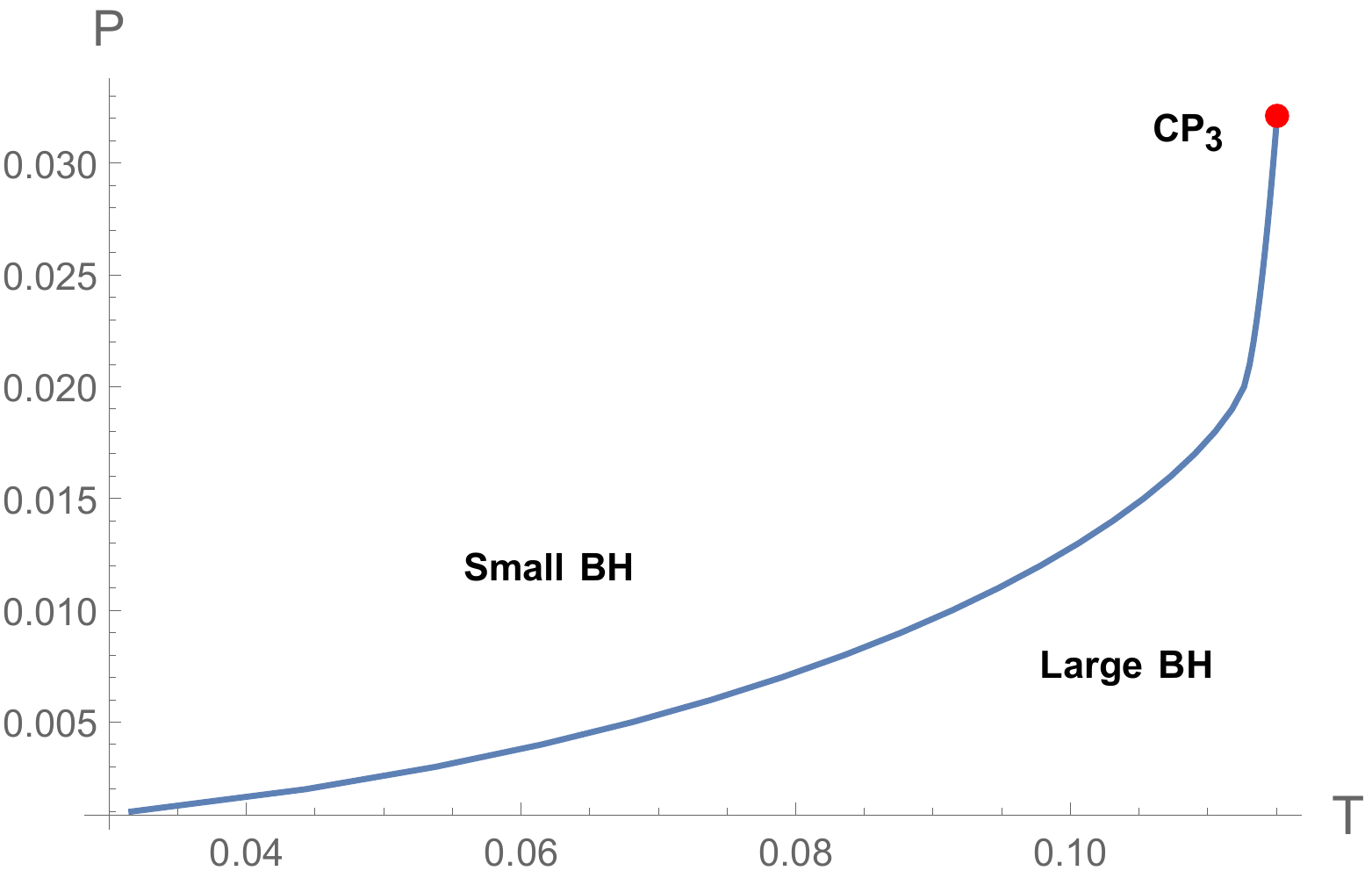}\label{Fig:ptplot_q15}}\hspace{0.5cm}	
		\subfloat[]{\includegraphics[width=2.9in]{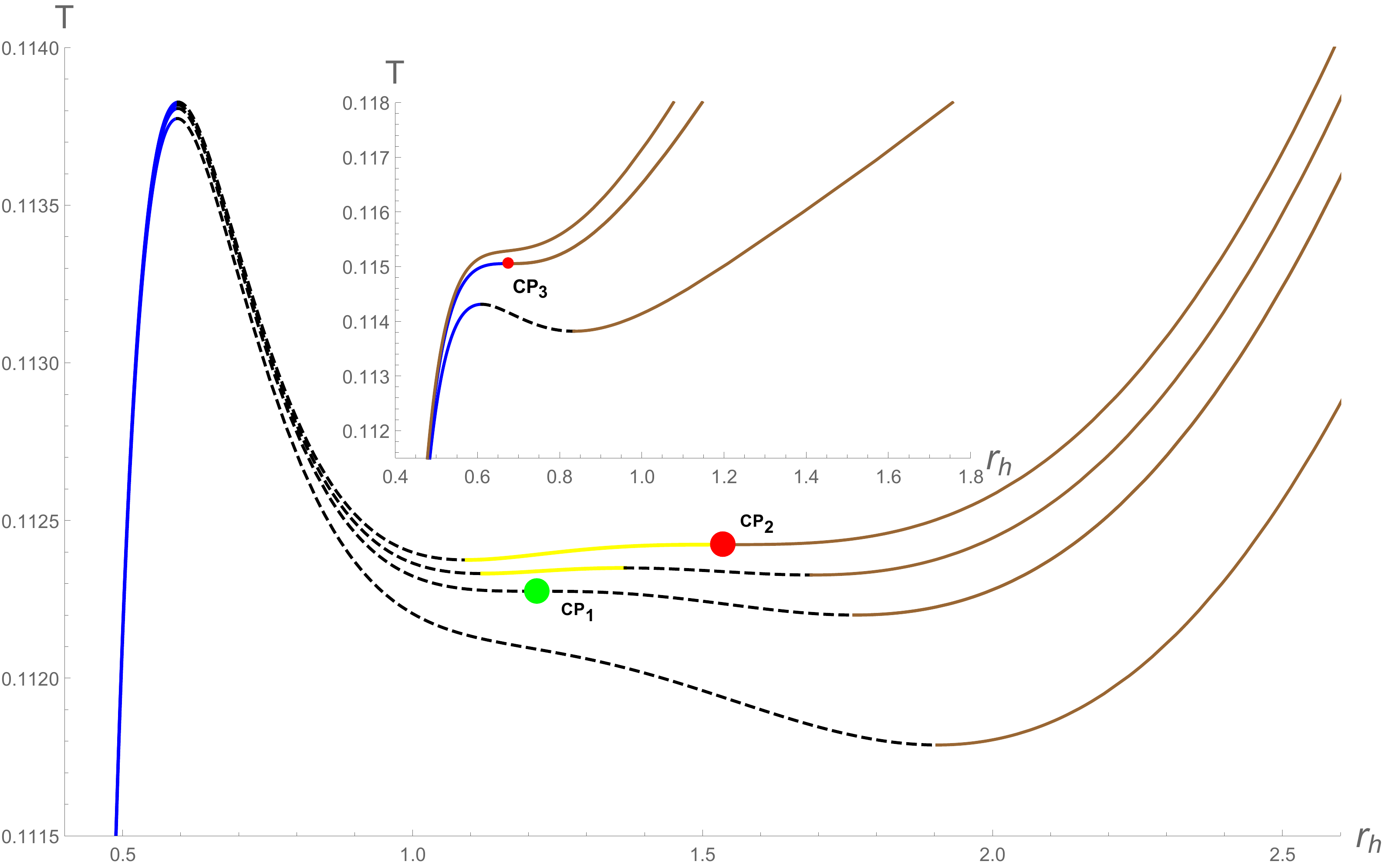}\label{Fig:trplot_q15}}				

		\caption{\footnotesize For case-2: (a) Phase diagram showing the first-order phase transitions near the conventional critical point $\text{CP}_3$.  (b)  $T$ as a function of $r_h$ for different pressures (Inset: for $P > P_{c2}.$), showing the appearance of new phases (stable or unstable) near the novel critical point $\text{CP}_1$ (green dot), and disappearance of phases near the conventional critical points $\text{CP}_2$, and $\text{CP}_3$ (red dots).  Dashed curves denote unstable black hole branches and solid curves denote stable black hole branches. Pressure of the isobars increases from bottom to top, where the critical pressures are $P_{c1}$, $P_{c2}$, $P_{c3}$ with $P_{c1} < P_{c2} < P_{c3}$.} 
	}
\end{figure}

\begin{figure}[h!]
		
	{\centering
	
		\subfloat[]{\includegraphics[width=2.9in]{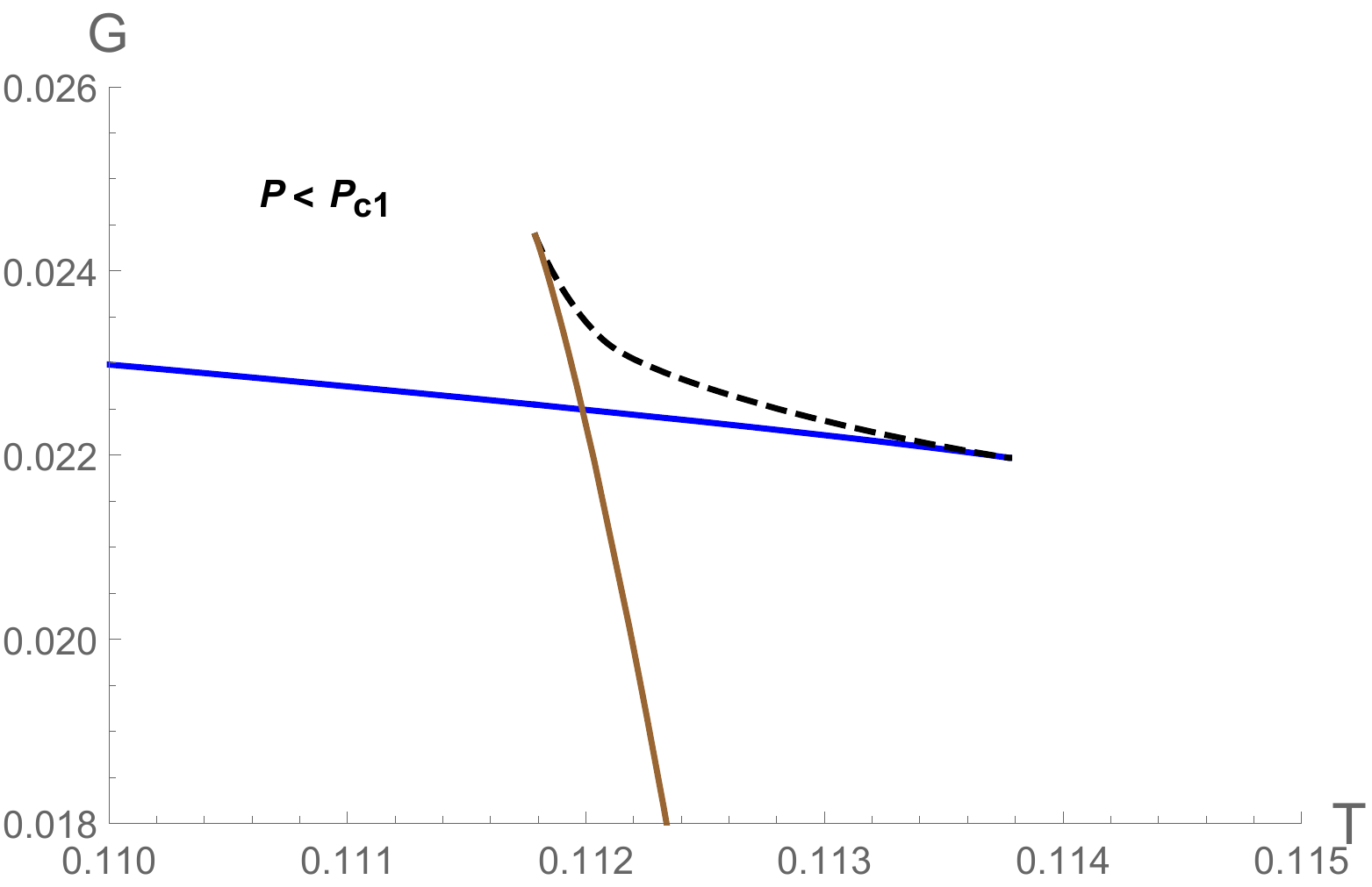}}\hspace{0.5cm}	
		\subfloat[]{\includegraphics[width=2.9in]{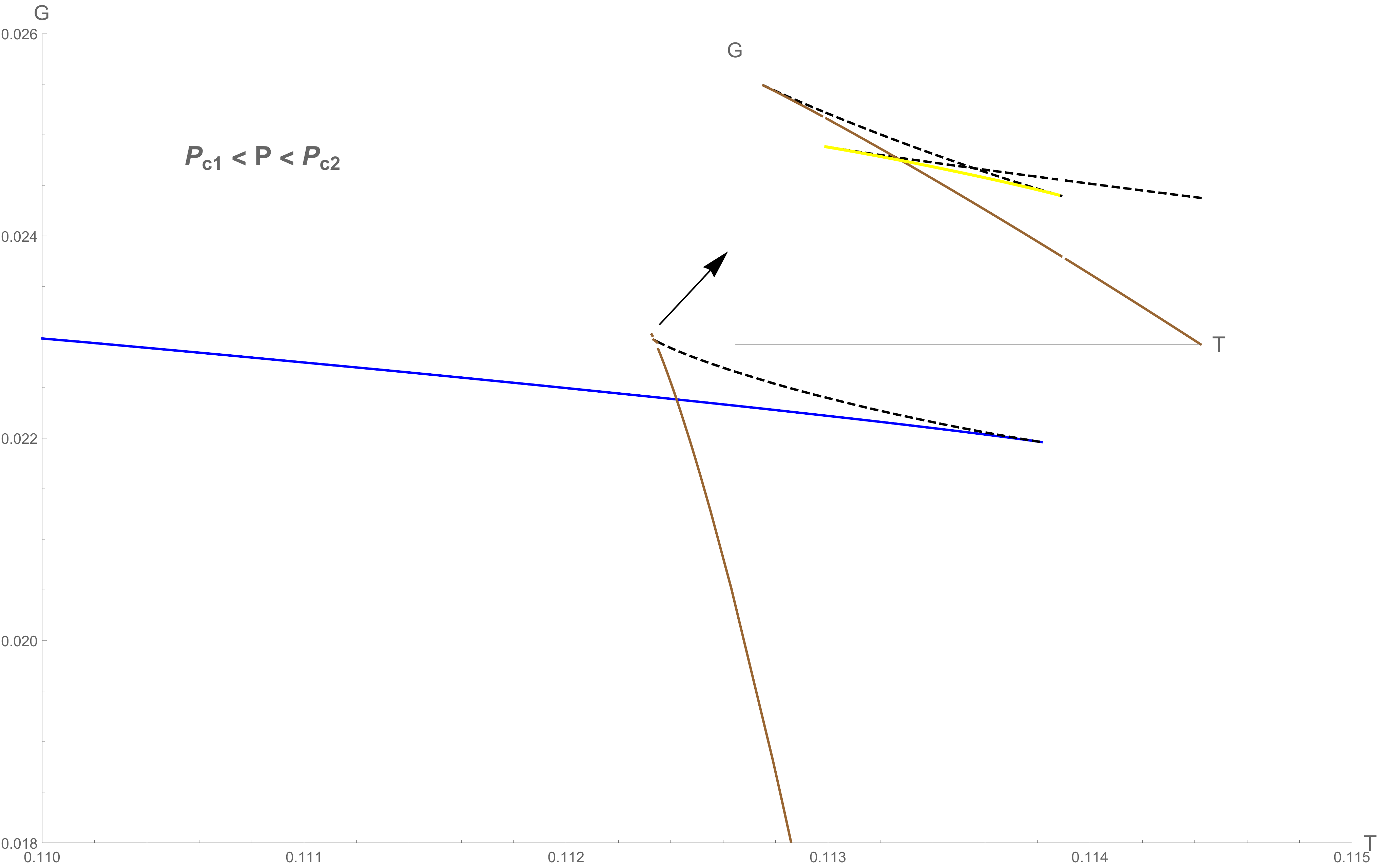}}\hspace{0.5cm}				
		\subfloat[]{\includegraphics[width=2.9in]{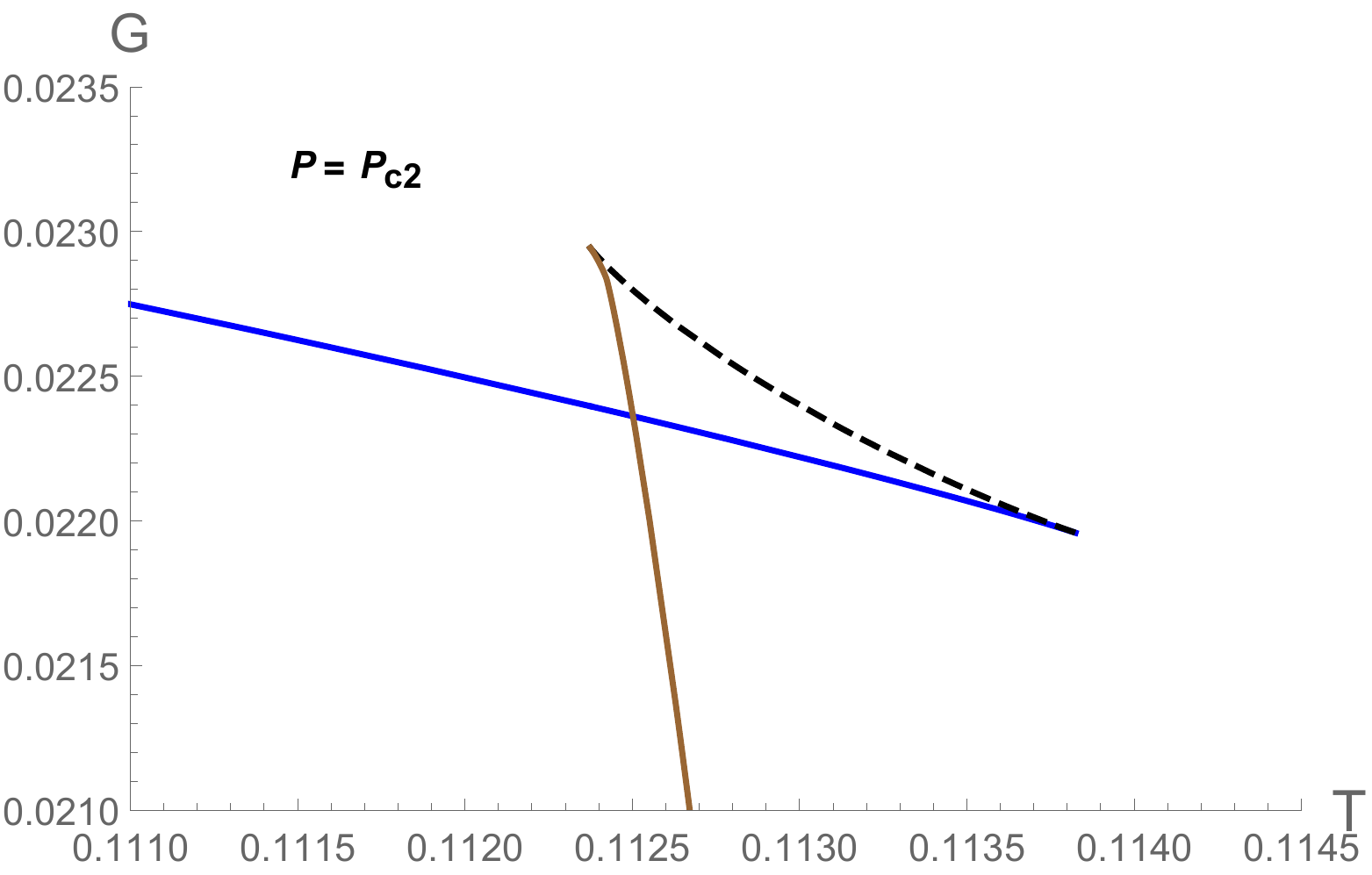}}\hspace{0.5cm}				
		\subfloat[]{\includegraphics[width=2.9in]{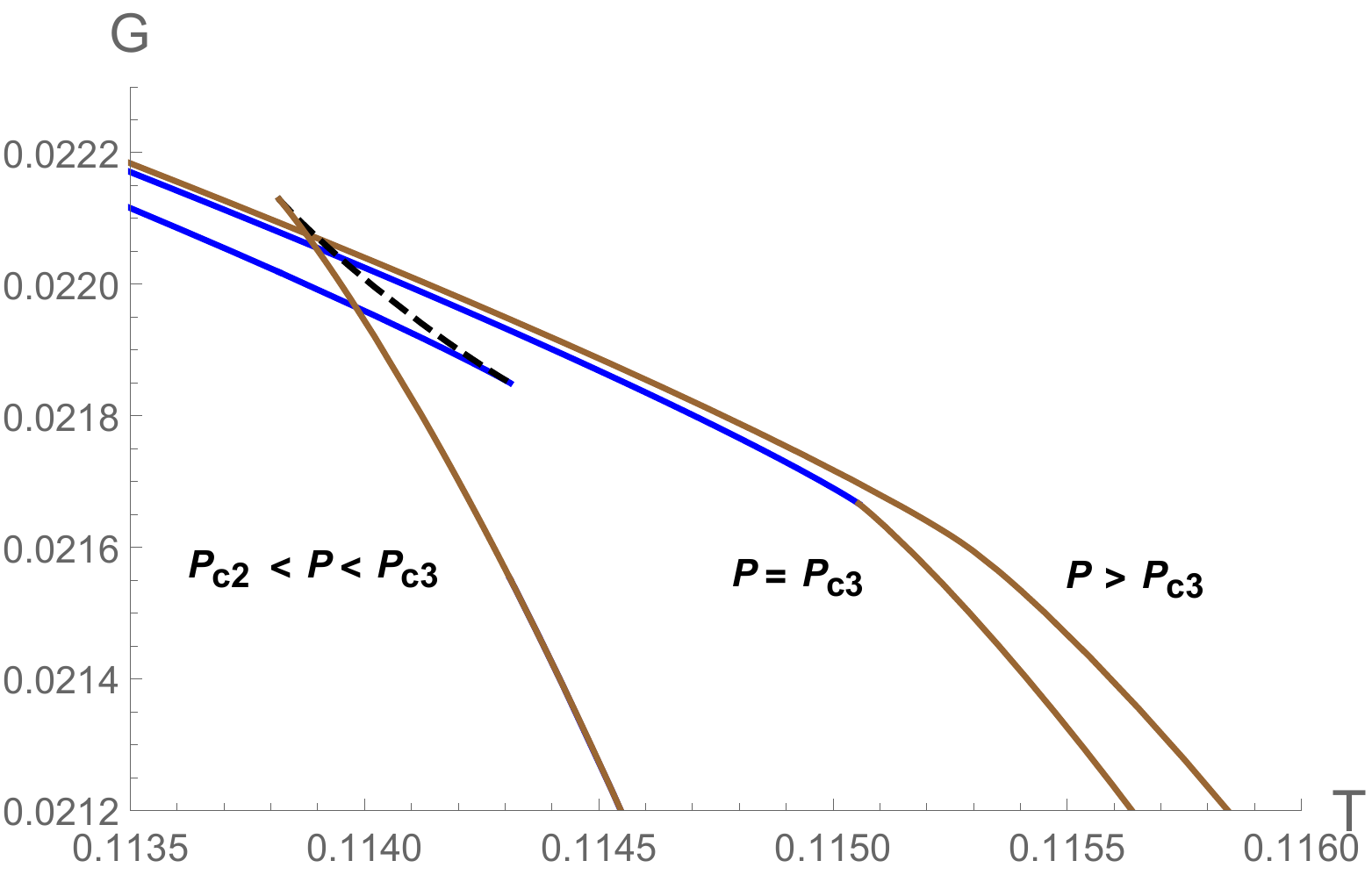}}
					
		
		\caption{\footnotesize For case-2: Behaviour of the Gibbs free energy as a function $T$ for different pressures,  showing the appearance of new (stable or unstable) phases near the novel critical point $\text{CP}_1$ (with pressure $P_{c1}$), and disappearance of  phases near the conventional critical points $\text{CP}_2$ ( with pressure $P_{c2}$), and $\text{CP}_3$ (with pressure $P_{c3}$), on increase of pressure. (a) $P=0.0192$, (b) $P=0.01968$,  (c) $P=P_{c2} = 0.01978$,  (d) $P=0.025, P_{c3}=0.0321, 0.034$.} 
	\label{fig:gtplots_q15}	}

\end{figure}

According to the proposal in~\cite{Wei:2021vdx}, we have two types of critical points from the topological classification. The conventional critical point (for which the topological charge is $Q_t = -1$), and the novel critical point (for which the topological charge is $Q_t = +1$). A further proposal of~\cite{Wei:2021vdx} is that,
the conventional critical point indicates the presence of first-order phase transitions near it, while the novel critical point cannot serve as an indicator of the presence
of a first-order phase transition. However, this notion of the presence or absence of first-order phase transitions based on the classification of the critical point as conventional or novel, does not seem to be true in general. This can be inferred, for example, from our results in case-2 (i.e., section~\ref{case:2}). 
\vskip 0.3cm \noindent
In case-2, we obtained  the topological charges of the three critical points $\text{CP}_1$, $\text{CP}_2$, and $\text{CP}_3$, as $Q_t|_{\text{CP}_1} = +1$ (novel), $Q_t|_{\text{CP}_2} = -1$ (conventional), and $Q_t|_{\text{CP}_3} = -1$ (conventional), respectively.
As shown in Fig.~\ref{Fig:ptplot_q15},  only the critical point $\text{CP}_3$ appears in the phase diagram, while the other two critical points do not appear, as they do not globally minimise the Gibbs free energy~\cite{Wei:2014hba,Frassino:2014pha}.
There exist a first-order phase transitions near the critical point $\text{CP}_3$ (conventional), and there are no first-order phase transitions near the critical point $\text{CP}_1$ (novel). However, even though the critical point $\text{CP}_2$ is a conventional one, there is no first-order phase transitions near it, which disaccords with the proposal in~\cite{Wei:2021vdx}. 
\vskip 0.3cm \noindent
We can resolve this disagreement, if we classify the critical points in the following way. As the pressure increases, the novel critical point is the one from which new phases (stable or unstable)  appear, whereas, the conventional point is the one at which the phases disappear, as can be seen from Figs.~\ref{Fig:trplot_q15}, and~\ref{fig:gtplots_q15}.
\vskip 0.3cm \noindent

\begin{figure}[h!]
	{\centering
		\subfloat[]{\includegraphics[width=2.9in]{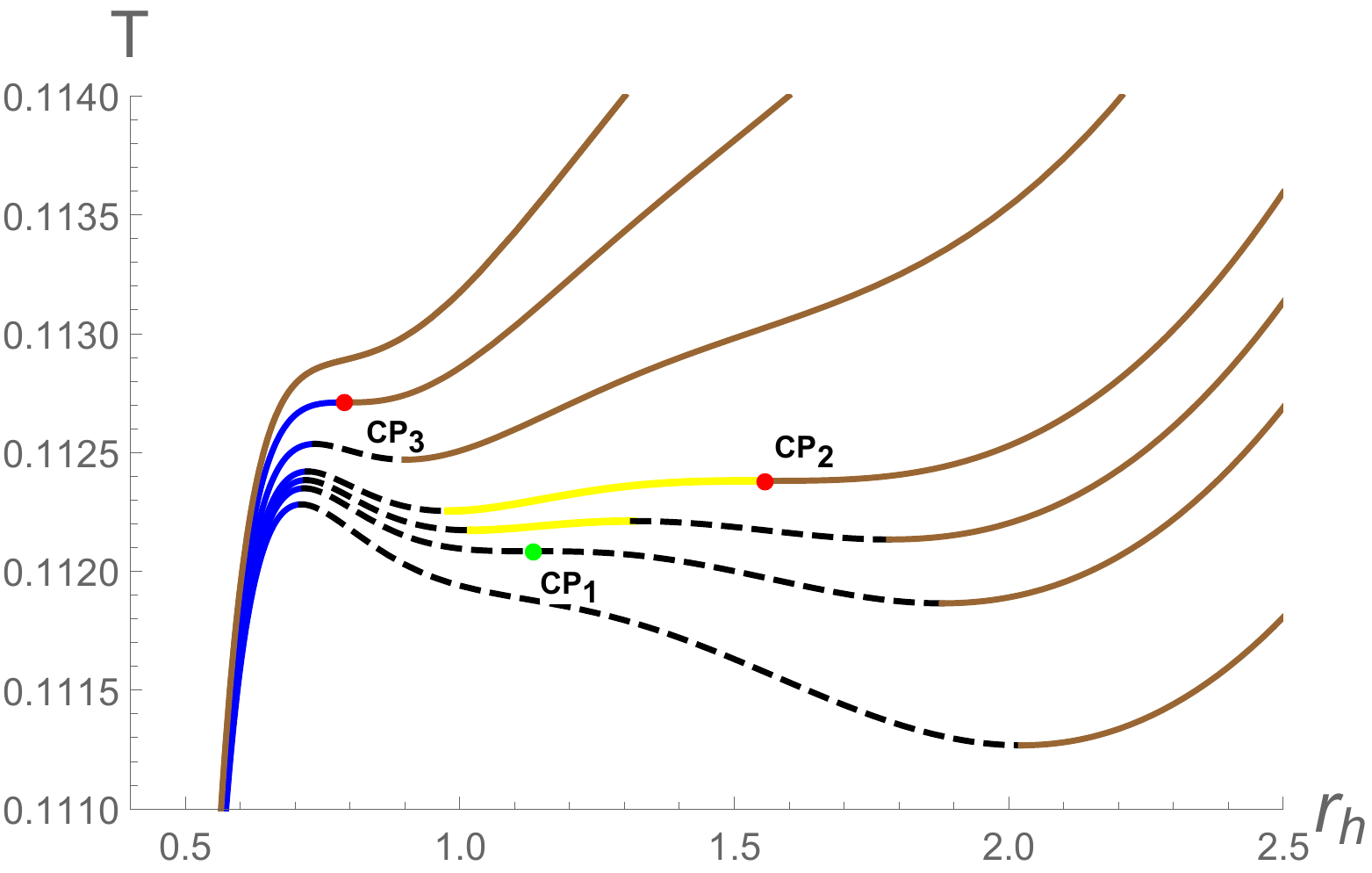}\label{Fig:trplot_q18}}\hspace{0.5cm}	
		\subfloat[]{\includegraphics[width=2.9in]{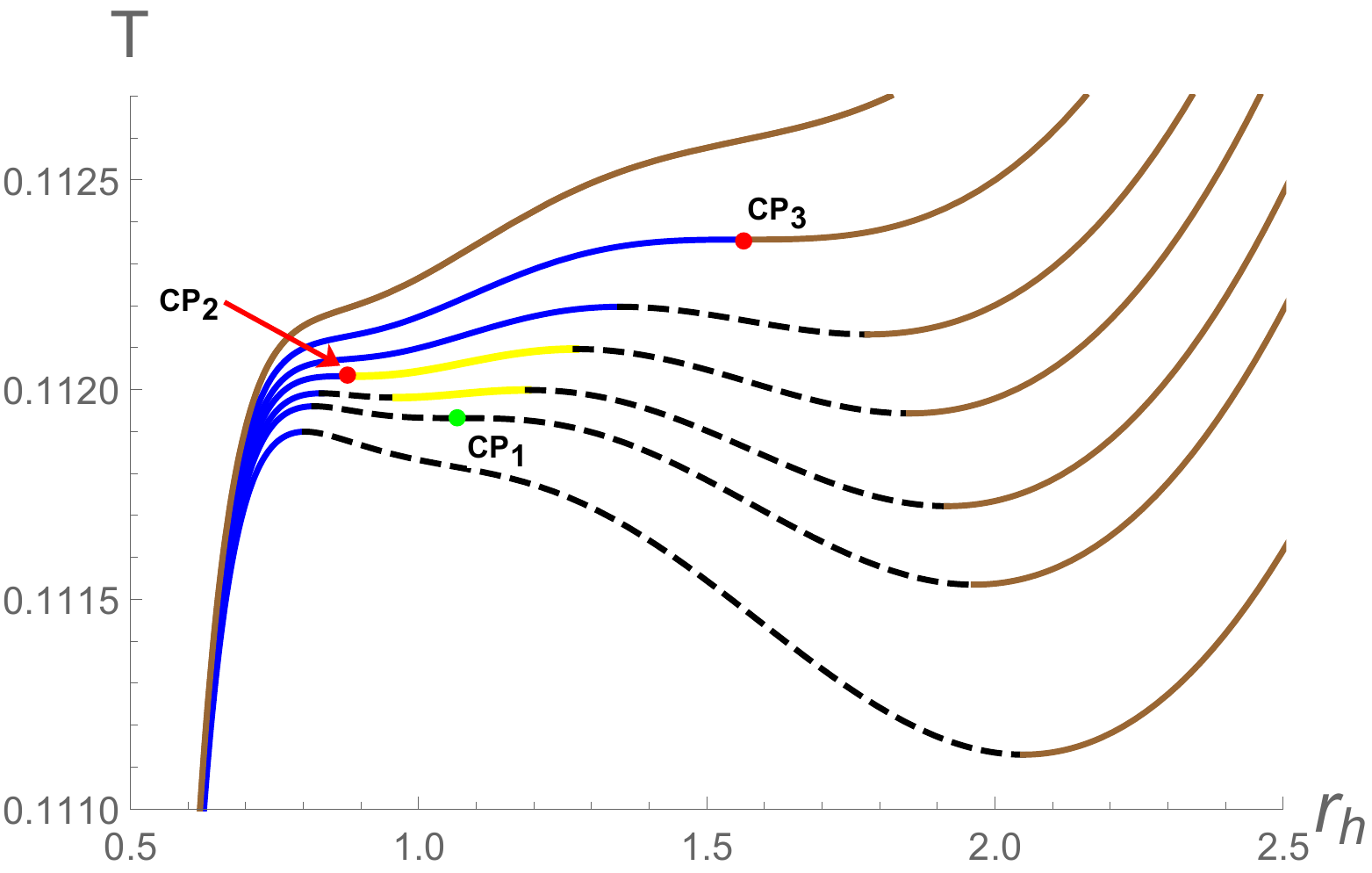}\label{Fig:trplot_q195}}				
		
		\caption{\footnotesize   $T$ as a function of $r_h$ for different pressures, showing the appearance of new phases (stable or unstable) near the novel critical point $\text{CP}_1$ (green dot), and disappearance of phases near the conventional critical points $\text{CP}_2$, and $\text{CP}_3$ (red dots).  Dashed curves denote unstable black hole branches and solid curves denote stable black hole branches. Pressure of the isobars increases from bottom to top, where the critical pressures are $P_{c1}$, $P_{c2}$, $P_{c3}$ with $P_{c1} < P_{c2} < P_{c3}$. (a) for case-3, (b) for case-4.} 
	}
\end{figure}   
\noindent
From the Fig.~\ref{Fig:trplot_q15},  we have three black hole phases  (small, unstable intermediate and large) when the pressure $P < P_{c1}$. We see the appearance of new phases from the critical point $\text{CP}_1$ (novel) when we just increase the pressure  $P > P_{c1}$, and we now have five phases (small, unstable intermediate, stable intermediate, unstable intermediate, large). On further increasing the pressure $P$ to $P_{c2}$, two phases among five are disappearing at the critical point $\text{CP}_2$ (conventional), and we are left with three phases only.
If we further increase the pressure $P$ to $P_{c3}$, two phases among three are disappearing  at the critical point $\text{CP}_3$ (conventional). This appearance/disappearance of phases near the critical points can be seen clearly from the Gibbs free energy behaviour shown in Fig.~\ref{fig:gtplots_q15}.
\vskip 0.3cm \noindent
For the cases 3 and 4, the appearance/disappearance of phases at the critical points can be seen from the Fig.~\ref{Fig:trplot_q18}, and~\ref{Fig:trplot_q195}, respectively.
It can be checked that our novel proposal of the appearance/disappearance of phases from the nature of critical points is also valid for the Born-Infeld case~\cite{Wei:2021vdx} and the third order Lovelock gravity presented in Appendix-(\ref{AppendixB}).

\section{Conclusions} \label{conclusions}
 In this paper, we considered six dimensional charged AdS black holes in Gauss-Bonnet gravity in the extended phase space, where the black holes exhibit a rich phase structure admitting multi critical points, depending on the range of the black hole charge $Q$.
 We computed the topological charges corresponding to the critical points by following Duan's topological current $\phi$-mapping theory and the proposal in~\cite{Wei:2021vdx}. Our findings are summarised below.
\vskip0.3cm \noindent
We first considered the case of charged AdS black hole by switching-off the Gauss-Bonnet (GB) coupling $\alpha$ (calculation presented in appendix-(\ref{AppendixA}). In this case, we have only one critical point for which the topological charge $Q_t$ is found to be $ Q_t|_{\text{CP}_1}=-1$, and thus this critical point is a conventional one, following the classification in~\cite{Wei:2021vdx}.
\vskip0.3cm \noindent
Next, we considered the effect of GB coupling on the topological charge. Depending on the range of the black hole charge $Q$, we now have three critical points for $ Q < Q_B $, and one critical point for $ Q > Q_B $ (where, $Q_B =0.2018$ at $\alpha = 1$.), and the corresponding topological charges are given in the Table~\ref{table:summary}.
Among the three critical points for $ Q<Q_B$,  we have one novel critical point and the other two are conventional critical points, while the critical point for $Q>Q_B$ is a conventional one. However, the total topological charge of the system is $-1$ independent of the range of the black hole charge $Q$.
\vskip 0.3cm \noindent
Therefore, we find that the topological charge of the charged AdS black hole system is ($Q_t = -1$) unaltered due to the Gauss-Bonnet coupling, unlike the Born-Infeld coupling~\cite{Wei:2021vdx}. 
Further, we found that as the pressure increases in phase space, new phases (stable or unstable) appear from the novel critical point, whereas the phases disappear at the conventional critical point. Thus, we name the novel critical point as the  phase creation point, and the conventional critical point as the phase annihilation point. This removes some ambiguity in the connection between existence/absence of first order phase transition and classification of critical points based on topology.
\begin{table}[h!]
	\centering{\
		\begin{tabular}{|c |c |c|c|c |c|c|}
			\hline \hline 
	Case & Number of  & Topological & Total Topological \\
		&  Critical Points &  Charge &  Charge $Q_t$ \\ \hline
			\multirow{3}{11em}{Case 1: $\alpha = 0$} &  &  &   \\ 
		                                        	& 1 & $Q_t|_{\text{CP}_1} = -1$ & $-1$  \\ 
			                                        &  &  &   \\ \hline
             \multirow{3}{11em}{Case 2: $ Q =0.15 < Q_C$} &  & $Q_t|_{\text{CP}_1} = +1$ &   \\ 
			                                        & 3 & $Q_t|_{\text{CP}_2} = -1$ & $-1$  \\ 
			                                        &  & $Q_t|_{\text{CP}_3} = -1$ &   \\ \hline
			 \multirow{3}{11em}{Case 3: $ Q_C < Q =0.18 < Q_D $} &  & $Q_t|_{\text{CP}_1} = +1$ &   \\ 
			                                         & 3 & $Q_t|_{\text{CP}_2} = -1$ & $-1$  \\ 
			                                         &  & $Q_t|_{\text{CP}_3} = -1$  &   \\ \hline
			 \multirow{3}{11em}{Case 4:  $ Q_D < Q =0.195 < Q_B $ } &  & $Q_t|_{\text{CP}_1} = +1$ &   \\ 
			                                         & 3 & $Q_t|_{\text{CP}_2} = -1$ & $-1$  \\ 
			                                         &  & $Q_t|_{\text{CP}_3} = -1$ &   \\ \hline
			 \multirow{3}{11em}{Case 5: $Q_B < Q =0.3$} &  &  &   \\ 
			                                           & 1 & $Q_t|_{\text{CP}_1} = -1$ & $-1$  \\ 
			                                           &  &  &   \\ \hline
		\end{tabular}
	\caption{Summary of the results.}
	\label{table:summary}}
\end{table}

\vskip 0.3cm \noindent
First observation is with regards to the parity of critical points. Black holes in two different systems with same parity (odd-odd or even-even type) of the number of critical points, have the total topological charge matching with the parity. Examples to support this view are charged AdS black hole (1 critical point) and charged GB AdS black hole (1 or 3 critical points), where the parity of the number of critical points (odd in both cases) is same irrespective of the sign of topological charge. On the other hand, in the case of charged black holes in AdS (1 critical point) and Born-Infeld AdS black hole (2 critical points), the parity of the number of critical points is different, i.e., odd in the first and even in the same case, respectively. Correspondingly, the total topological charge in these cases matches the parity of number of critical points, i.e., odd (-1) for charged black holes and even (0) with Born-Infeld corrections. This trend continues for black holes in the third order Lovelock theories (appendix-\ref{AppendixB}), where for odd (even) number of critical points the total topological charge is an odd (even) number. Of course, our study is preliminary and more examples are required to settle these claims.
\vskip0.3cm \noindent
An important result concerns the observations in~\cite{Wei:2021vdx}, where for black holes in the Einstein-Maxwell system and their counterparts in the Einstein-Born-Infeld system, the critical points were shown to belong to different topological classes, as their topological charges are different. On the other hand, as we found in this work, that the charged black holes in the Gauss-Bonnet gravity belong to the same topological class as their counterparts in the Einstein-Maxwell system. As shown in appendix-(\ref{AppendixB}), for black holes in Lovelock theories of gravity up to third order the topological class of critical points remains unaltered. Naively, it appears that the higher derivative curvature corrections in the gravity sector do not change the topological class of black hole critical points, where as the gauge corrections do. There might be a deeper reason for this which requires further study, possibly by studying black holes in theories with both higher curvature gauge and gravity corrections side by side in the action.  It would be interesting to explore this issue further from topology point of view, as it may teach us something new about the phase structure of black holes, which has been missed from standard thermodynamic treatments. There is no doubt that, the above conjectures require more scrutiny in systems having multiple critical points and with other gauge/gravity corrections, possibly involving rotating black hole solutions~\cite{Frassino:2014pha,Hendi:2017mfu,Zhang:2017lhl,Momennia:2021ktx,Wei:2015ana,Li:2022vcd}. Further, it would be nice to have a mechanism to study the topological properties associated with the triple points in black holes as well~\cite{Wei:2014hba,Altamirano:2013uqa,Frassino:2014pha}.

\renewcommand{\thesection}{\Alph{section}}
\appendix
\section*{Appendices}

\section{Case-1 $(\alpha=0)$: Topology of Einstein-Maxwell system in six dimensions} \label{AppendixA}

\begin{figure}[h!]
	{\centering
		\subfloat[]{\includegraphics[width=2.7in]{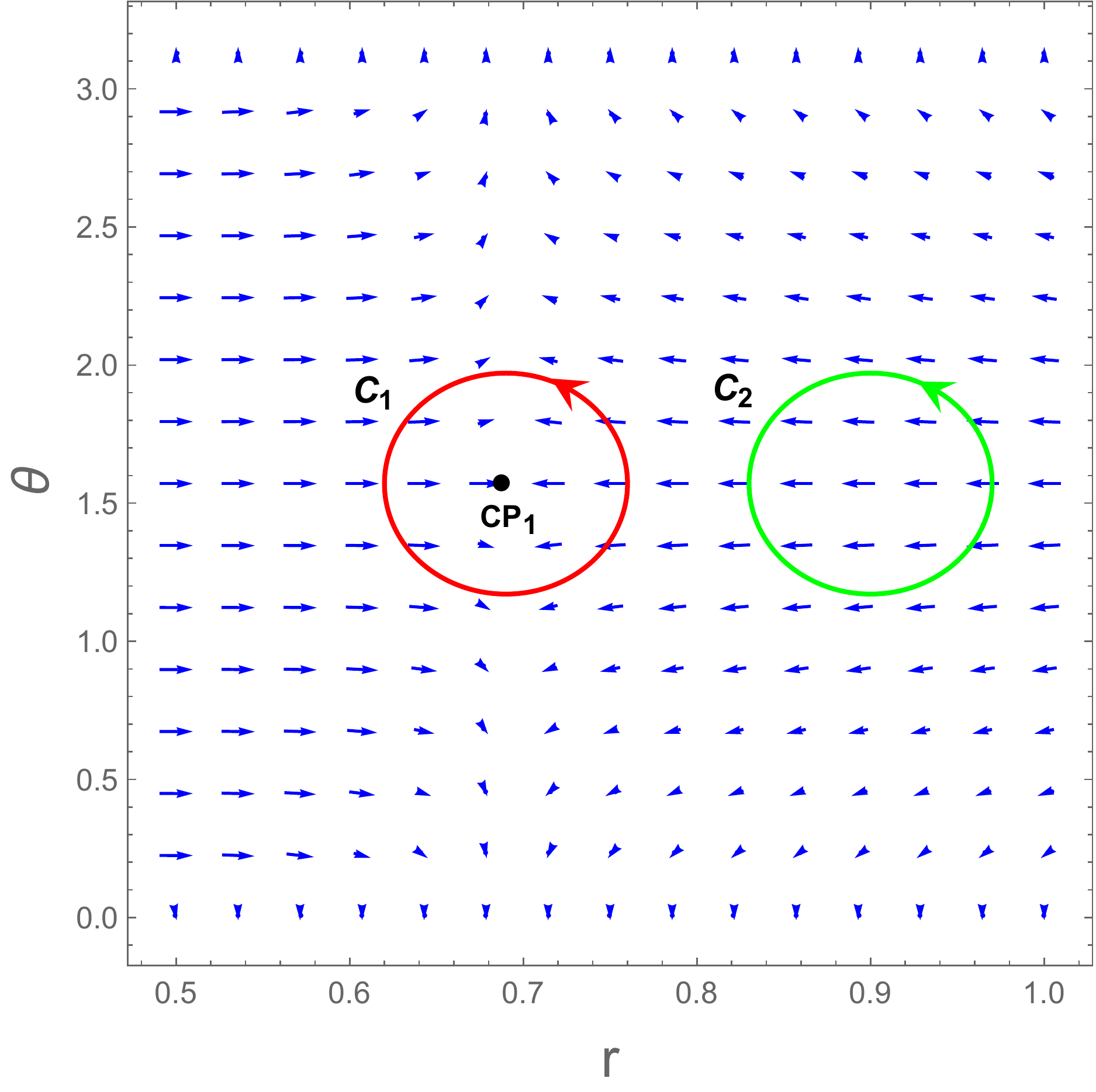}\label{Fig:vecplot_max}}\hspace{0.5cm}	
		\subfloat[]{\includegraphics[width=2.7in]{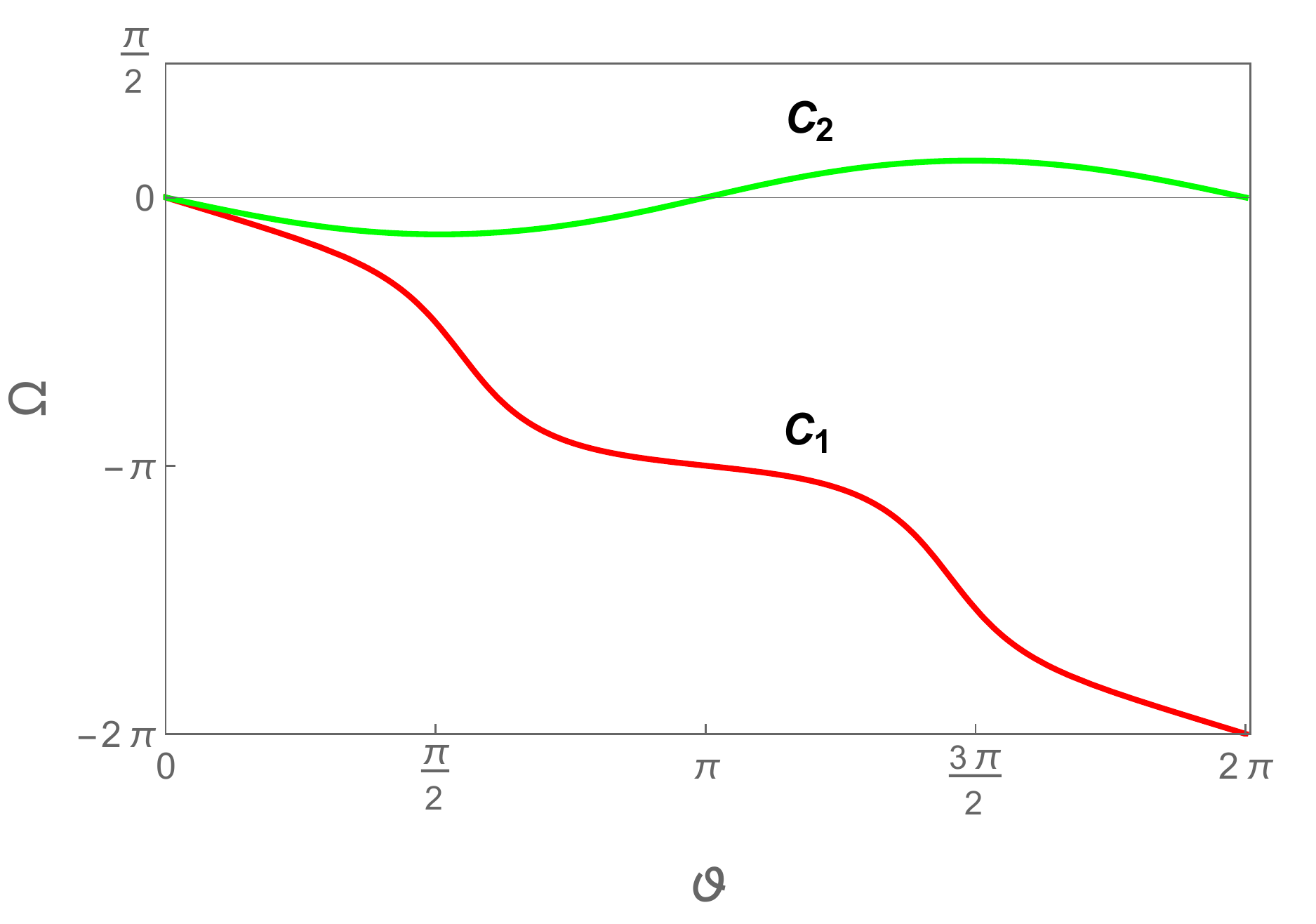}\label{Fig:omegaplot_max}}				
		
		\caption{\footnotesize For case-1: (a) The blue arrows represent the vector field $n$ on a portion of the $\theta-r$ plane. The critical point $\text{CP}_1$ is located at $(r,\theta)=(r_c, \frac{\pi}{2})$ marked with a black dot.
			The  contour $C_1$  encloses the critical point, while the contour $C_2$ does not. (b) $\Omega$ vs $\vartheta$ for contours $C_1$ (red curve) and $C_2$ (green curve). (Here, we set the charge $Q=0.15$).} 
	}
\end{figure}

The aim here is to compute the topological charge $Q_t$ associated with the Einstein-Maxwell system by switching-off the Gauss-Bonnet parameter  (i.e., $\alpha = 0$) in eqn.~\eqref{eq:GB_eq of st} and this will help validate our results when the charge parameter is introduced in section-(\ref{topologygb}). This system has only one critical point, showing the van der Waals type small/large black hole phase transitions, given by~\cite{Gunasekaran:2012dq,Wei:2014hba}
\begin{equation}
T_c = \frac{9}{7\pi r_c}, \quad P_c = \frac{9}{16\pi r_c^2}, \quad r_c= \Big(\frac{14 Q^2}{3}\Big)^{1/6}.
\end{equation}
For this case, the vector field $n$ is plotted in Fig.~\ref{Fig:vecplot_max}, where it shows the critical point at $(r,\theta) = (r_c, \frac{\pi}{2})$. We then construct two contours $C_1$
and $C_2$ such that the contour $C_1$ encloses the critical point, while the contour $C_2$ does not (see Table~\ref{table:coeff} for parametric coefficients of the contours.).
\vskip 0.3cm \noindent
The deflection angle $\Omega(\vartheta)$  (using eqn.~\eqref{eq:deflection}) of the vector field $\phi$ along the contours $C_1$ and $C_2$ is plotted in Fig.~\ref{Fig:omegaplot_max}. It shows the following features.  For contour $C_1$, it decreases and reaches $\Omega (2\pi) = -2\pi$, while for the contour $C_2$, it decreases first and then increases, before finally vanishing i.e., $\Omega (2\pi) = 0$.
Therefore, the topological charge $Q_t = \frac{1}{2\pi} \Omega (2\pi)$  for the contour  $C_1$ is $-1$, while for the contour $C_2$ it is zero. These results are consistent with our expectations that,  if a given contour encloses the critical point then its topological charge is non-zero.  
Therefore, the topological charge corresponding to the critical point $\text{CP}_1$ is $Q_t|_{\text{CP}_1} = -1$. According to the classification of the critical points from topology proposed in~\cite{Wei:2021vdx}, this critical point is a conventional critical point. Since there exists only one critical point, the total topological charge of the Einstein-Maxwell system would be $Q_t = -1.$

\newpage
\section{Topology of critical points in third order Lovelock Gravity} \label{AppendixB}
\begin{figure}[h!]
     	{\centering
     		\subfloat[]{\includegraphics[width=2.9in]{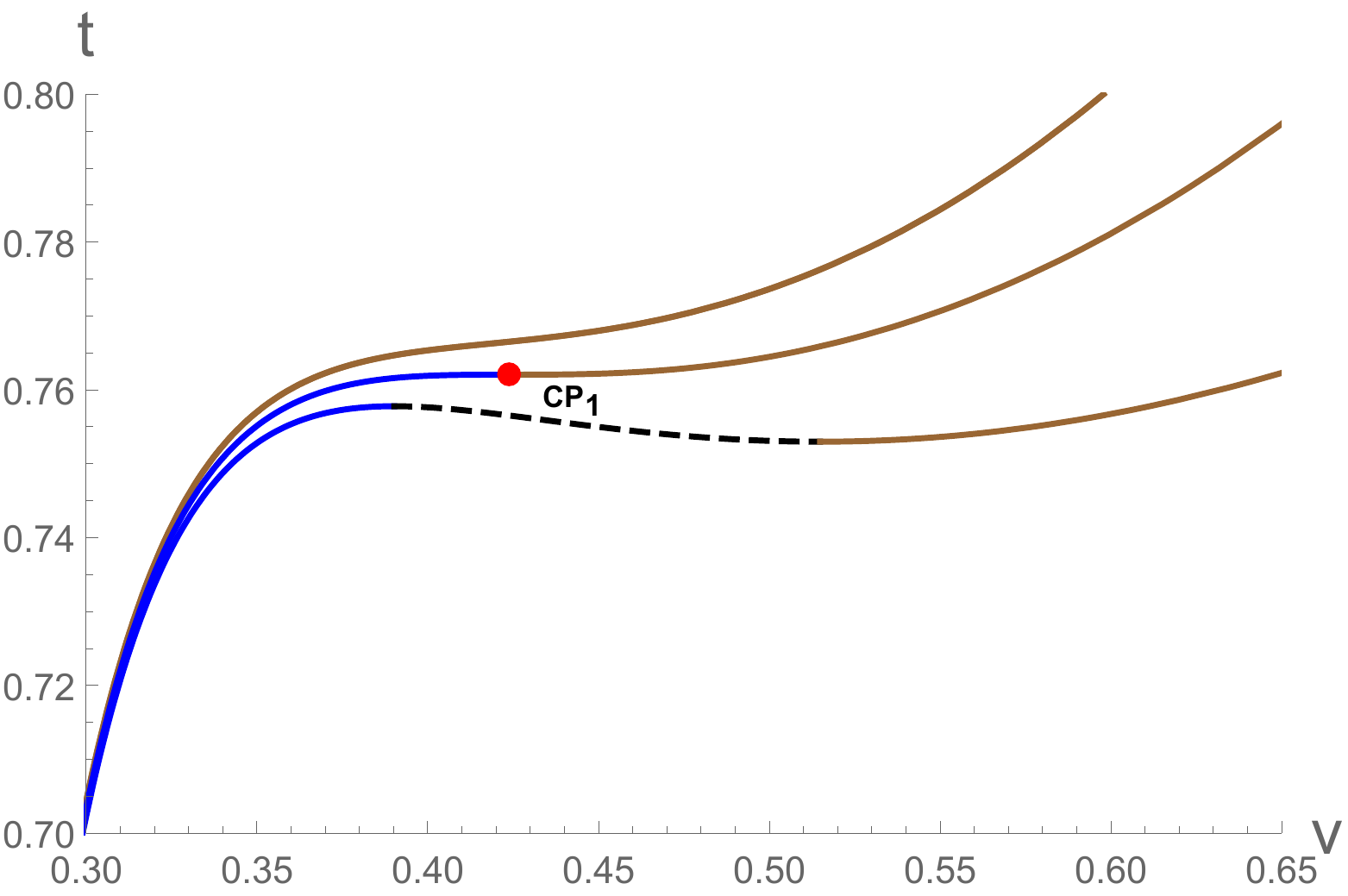}\label{Fig:tvplot_lov_1cp}}\hspace{0.2cm}	
     		\subfloat[]{\includegraphics[width=2.9in]{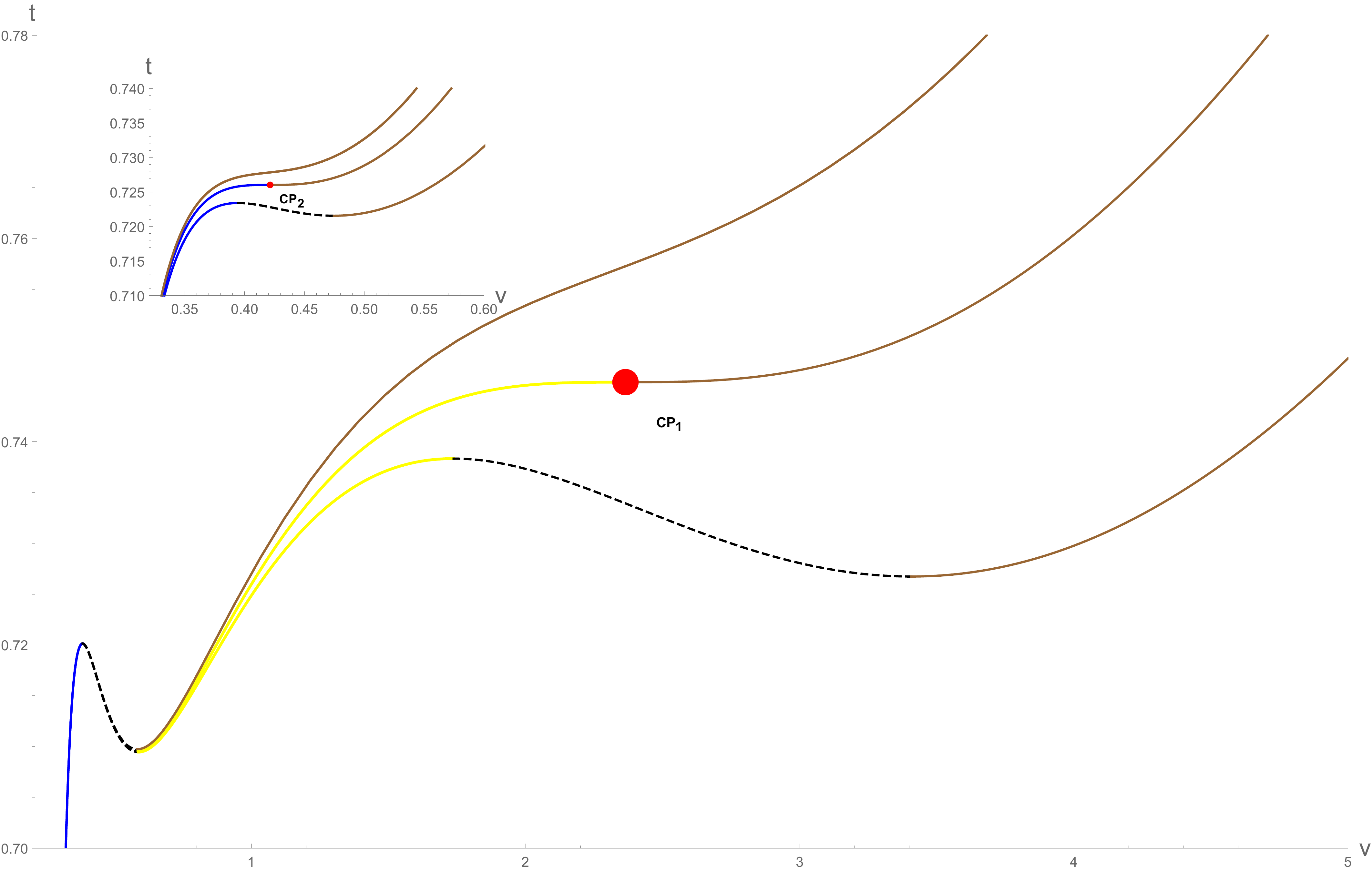}\label{Fig:tvplot_lov_2cp}}\hspace{0.5cm}
     					
     		
     		\caption{\footnotesize For Lovelock gravity:  $t$ as a function of $v$ for different pressures. The critical points shown in red colour are phase annihilation points.  Dashed curves denote unstable black hole branches and solid curves denote stable black hole branches. Pressure of the isobars increases from bottom to top.  (a) with one critical point, (b) with two critical points (Inset: for  $p>p_{c1}$).} 
     	}
\end{figure}

\begin{figure}[h!]
     	{\centering
     		
		\includegraphics[width=5.8in]{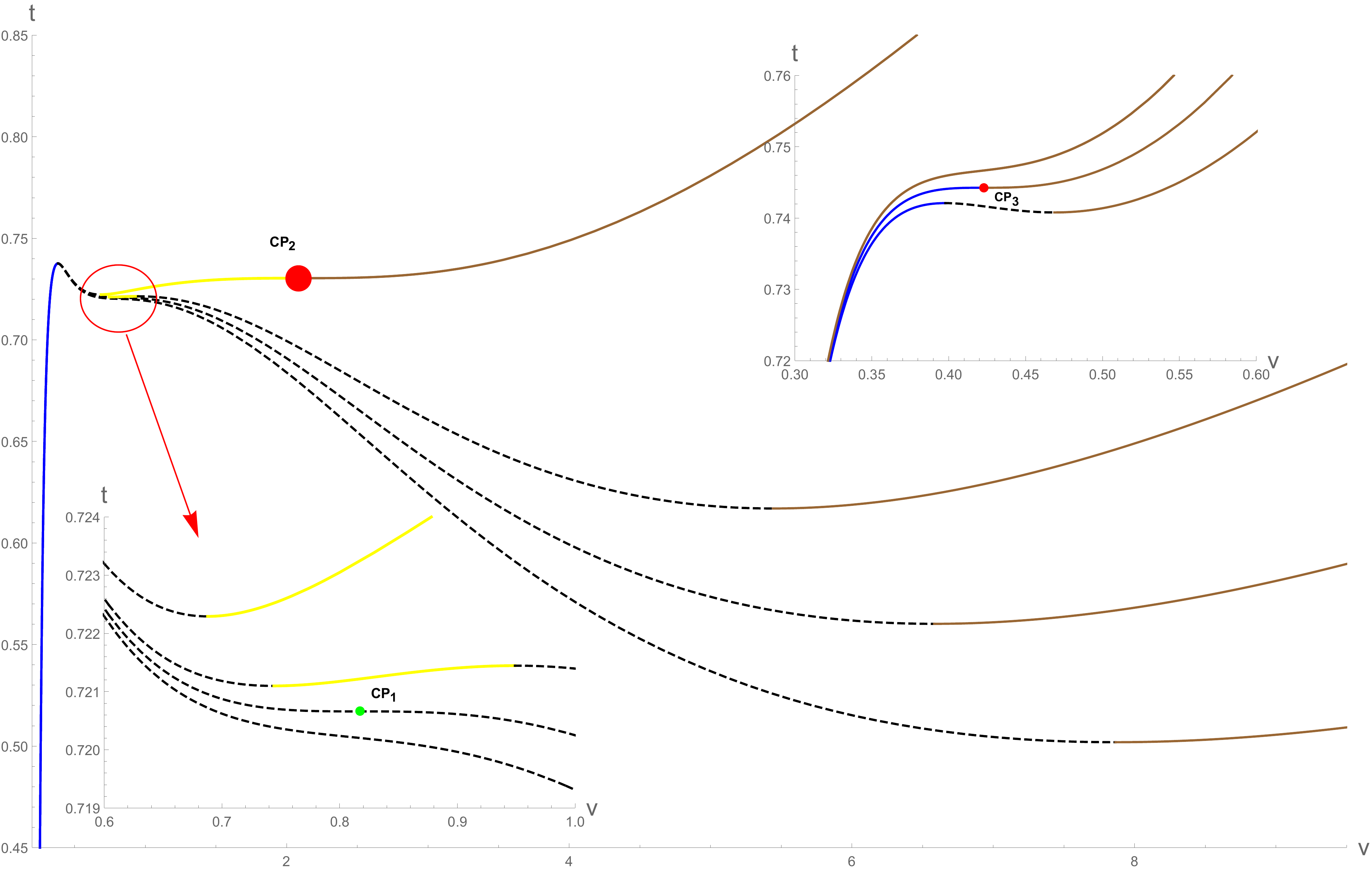}
				
		
     		\caption{\footnotesize For Lovelock gravity:  $t$ as a function of $v$ for different pressures. The critical points shown in red (green) colour are phase annihilation (phase creation) points.  Dashed curves denote unstable black hole branches and solid curves denote stable black hole branches. Pressure of the isobars increases from bottom to top, with three critical points (Insets: for  $p>p_{c2}$).} \label{Fig:tvplot_lov_3cp}	
     	}
     \end{figure}

            \begin{table}  
   \centering{
    \begin{tabular}{|c|c|c|c|c|c|c|}
    	\hline \hline
  Parameters & 	\multicolumn{2}{|c|}{Number of}   & \multicolumn{3}{|c|}{Critical Point (CP)}    \\ \cline{4-6}
    $(q, \alpha)$  & 	\multicolumn{2}{|c|}{Critical Points}  & $\text{CP}_1$ & $\text{CP}_2$ & $\text{CP}_3$ \\ \hline
     &  	\multirow{3}{3em} { 2}  & $v_c$ & 2.36519 & 0.4213 &  - \\ 
 $(0.01, 2.6)$   &  	& $p_c$ & 0.090224 & 2.46059 & -  \\ 
     &  	& $t_c$ & 0.745863 & 0.72603 & - \\ \hline
     &  	\multirow{3}{3em} { 3} & $v_c$ & 0.8172 & 2.0874 & 0.423 \\ 
    $(0.01,2.8)$  & 	& $p_c$ & 0.039  & 0.08996 & 2.7776 \\ 
      & 	& $t_c$ &  0.7206 & 0.7304 & 0.7442 \\ \hline
     &  	\multirow{3}{3em} { 1} & $v_c$ & 0.4241 & - & -  \\ 
    $(0.01,3)$  & 	& $p_c$ & 3.1603 & - & - \\ 
      & 	& $t_c$ & 0.7621 & - & -\\ 	\hline\hline   
    \end{tabular}
    \caption{Critical values for Lovelock gravity.}
    \label{table:cri for love}}
     \end{table}
Following the computation of topological charge and analysis in sections-(\ref{6GB}) and (\ref{naturegb}), we found that  in the case of Gauss-Bonnet black holes in AdS, the nature of critical points need to be understood as phase creation and phase annihilation points. Here, our aim is to give a prefatory computation that the arguments in section-(\ref{naturegb}) continue to hold in Lovelock theories of gravity as well. As an example, we consider the third-order Lovelock gravity but generalisation to higher order should also be possible. Following~\cite{Frassino:2014pha}, it is known that the third-order Lovelock coupling to Einstein-Maxwell system shows the standard van der Waals behaviour with one critical point in seven dimensions, and thus its (total) topological charge would be $-1$ (conventional critical point). However, this system in higher dimensions ($\geq 8$) exhibits the rich phase structure with multi critical points~\cite{Frassino:2014pha}. In eight dimensions, this system can have up to three critical points (depending on the charge $q$ and Lovelock coupling parameter $\alpha$), having the equation of state given by~\cite{Frassino:2014pha}:
 \begin{equation}
 p= \frac{t}{v} -\frac{15}{2\pi v^2} + \frac{2\alpha t}{v^3} -\frac{9\alpha}{2\pi v^4} + \frac{3t}{v^5}-\frac{3}{2\pi v^6} + \frac{q^2}{v^{12}},
 \end{equation}
 where, $v$ is the specific volume. The corresponding phase diagrams with one, two, and three critical points  (see Table~\ref{table:cri for love} for critical values) are as shown in Figs.~\ref{Fig:tvplot_lov_1cp},~\ref{Fig:tvplot_lov_2cp}, and~\ref{Fig:tvplot_lov_3cp}, respectively.

\vskip 0.3cm \noindent
 In Figures~\ref{Fig:tvplot_lov_1cp},~\ref{Fig:tvplot_lov_2cp}, and ~\ref{Fig:tvplot_lov_3cp}, the critical points shown in red colour correspond to the phase annihilation points and thus their topological charge is  $-1$. Where as, the critical point shown in green colour corresponds to the phase creation point and hence its topological charge is $+1$. 
  Therefore, the total topological charge of the Einstein-Maxwell system with Lovelock coupling would be $-1 \text{(with one or three critical points)}, \text{or}, -2\text{(with two critical points)}$. Thus, similar to cases of Gauss-Bonnet theories discussed in section-(\ref{naturegb}), we can advance the argument that the Lovelock coupling also does not alter the topological class of the critical points for black holes in Einstein-Maxwell system.	

\section*{Acknowledgements}
One of us (C.B.) thanks the DST (SERB), Government of India, for financial support through the Mathematical Research Impact Centric Support (MATRICS) grant no. MTR/2020/000135. We thank the anonymous referees for helpful suggestions which improved the manuscript.
\bibliographystyle{apsrev4-1}
\bibliography{topology_gb}
\end{document}